\documentclass[10pt]{article}
\usepackage{amsmath,amscd,amssymb,graphicx,latexsym,multicol,rotating,overpic}
\usepackage[T1]{fontenc}
\usepackage{color}

\newcommand\Rey{\mbox{\textit{Re}}}  % Reynolds number
\newcommand\Str{\mbox{\textit{St}}}
\newsavebox{\astrutbox}
\sbox{\astrutbox}{\rule[-5pt]{0pt}{20pt}}

\newcommand{\mA}{\boldsymbol{A}}
\newcommand{\mB}{\boldsymbol{B}}
\newcommand{\mC}{\boldsymbol{C}}
\newcommand{\mD}{\boldsymbol{D}}

\newcommand{\bsx}{\boldsymbol{x}}
\newcommand{\bsy}{\boldsymbol{y}}
\newcommand{\bsu}{\boldsymbol{u}}
\newcommand{\bs}[1]{\boldsymbol{#1}}
\newcommand{\m}[1]{\boldsymbol{#1}}

\newcommand{\bx}{\bsx}
\newcommand{\by}{\bsy}
\newcommand{\bu}{\bsu}

\title{\LARGE{\textbf{State-space model identification and feedback control \\of unsteady aerodynamic forces}}}
\newcommand*\samethanks[1][\value{footnote}]{\footnotemark[#1]}

\author{Steven L. Brunton$^\ddagger$\thanks{Department of Applied Mathematics, University of Washington, Seattle, WA. 98195-2420.  $^\ddagger$ ({sbrunton@uw.edu}). Questions, comments, or corrections to this document may be directed to that email address.} \and Scott T.~M.~Dawson\thanks{Mechanical and Aerospace Engineering, Princeton University, Princeton, NJ 08544 } \and Clarence W.~Rowley\samethanks[2]}
\date{}
    
\begin{document}
\maketitle

\begin{abstract}
Unsteady aerodynamic models are necessary to accurately simulate forces and develop feedback controllers for wings in agile motion; however, these models are often high dimensional or incompatible with modern control techniques.  Recently, reduced-order unsteady aerodynamic models have been developed for a pitching and plunging airfoil by linearizing the discretized Navier-Stokes equation with lift-force output.  In this work, we extend these reduced-order models to include multiple inputs (pitch, plunge, and surge) and explicit parameterization by the pitch-axis location, inspired by Theodorsen's model.  Next, we investigate the na\"{\i}ve application of system identification techniques to input--output data and the resulting pitfalls, such as unstable or inaccurate models.  Finally, robust feedback controllers are constructed based on these low-dimensional state-space models for simulations of a rigid flat plate at Reynolds number 100.  Various controllers are implemented for models linearized at base angles of attack $\alpha_0=0^\circ, \alpha_0=10^\circ$, and $\alpha_0=20^\circ$.  The resulting control laws are able to track an aggressive reference lift trajectory while attenuating sensor noise and compensating for strong nonlinearities. 
 \end{abstract}
 

%%%%%%%%%%%%%%%%%%%%
%% INTRODUCTION
%%%%%%%%%%%%%%%%%%%%
\section{Introduction}
Time-varying fluid flows are ubiquitous in modern engineering and in the life sciences, and controlling the corresponding unsteady aerodynamic forces and moments poses both a challenge and an opportunity.  Biological propulsion illustrates the potential utilization of unsteady forces for engineering design~\cite{Daniel:1984,Allen:2001,clark:2006,Dabiri:2009}.  It is observed that birds, bats, insects, and fish routinely exploit unsteady fluid phenomena to improve their propulsive efficiency, maximize thrust and lift, and increase maneuverability~\cite{birch:01buglev,Combes:2001,sane:03bug,wang:2005,Wu:2011,Shelley:2011}.  They achieve this performance with robustness to external factors, such as gust disturbances and weather, rapid changes in flight conditions, and even gross bodily harm.  At the same time, they do so with fixed actuators (wing muscles) and a limited number of noisy, distributed sensors throughout the body.  As uninhabited aerial vehicles (UAVs) become smaller and lighter, robust unsteady aerodynamic control will become increasingly important during agile maneuvers and gust disturbances.  

%%%%%%%%%%%%%%%%%%%%
%% NOMENCLATURE
%%%%%%%%%%%%%%%%%%%%
\begin{figure*}[t]
\framebox{
\noindent\begin{minipage}[h]{.53\columnwidth}
\section*{Nomenclature}
\begin{tabular}{ll}
$(\mA,\mB,\mC)$ & State-space model for transient lift\\
$(\mA,\mB,\mC)_r$ & Reduced-order model of order $r$\\
$a$ & Pitch axis with respect to $1/2-$chord\\
$b$ & Curvature parameter for step-up maneuvers\\
$c$ & Chord length of plate\\
$C_L$ & Lift coefficient [$C_L\triangleq{2L}/{\rho U_{\infty}^2c}$]\\
$C_{\alpha}$ & Lift coefficient slope in $\alpha$\\
$e$ & Noisy error signal \\ 
$f$ & Frequency of maneuver [Hz]\\
$G(s)$ & Transfer function for transient lift \\ 
$G_a$ & Actuator model\\
$g$ & Horizontal position of plate\\
$\mathcal{H}_i$ & $i$-th Markov parameter\\% [$\mathcal{H}_i\triangleq CA^{i-1}B$] \\
$h$ & Vertical position of plate\\
$k$ & Reduced frequency [$k\triangleq{\pi f c}/{U_{\infty}}$]\\
$L$ & Lift force\\
$L_d$ & Desired loop shape\\
$\mathcal{L}$ & Laplace transform\\
$M$ & Amplitude of motion\\
$n$ & Sensor noise
\end{tabular}
\end{minipage}
\begin{minipage}[h]{.43\columnwidth}
\begin{tabular}{ll}
$\text{Re}$ & Reynolds number [$\text{Re}\triangleq  cU_{\infty}/\nu$]\\
$r$ & Reduced-order model order\\
$r_L$ & Reference lift\\
$\text{St}$ & Strouhal number [$\text{St}\triangleq fA/U_{\infty}$]\\
$s$ & Laplace variable (dimensionless)\\
$t$ & Time (dimensional)\\
$\boldsymbol{U}$ & Vector of input motion\\
$U_{\infty}$ & Free stream velocity\\
$\bu$ & Input to state-space model\\
$\bx$ & State of state-space model\\
$\boldsymbol{Y}$ & Vector of measurements\\
$\by$ & Output of state-space model\\
$\alpha$ & Angle of attack of plate\\
$\alpha_e$ & Effective angle of attack\\
$\alpha_0$ & Base angle of attack\\
$\Delta t_c$ & Coarse time-step\\
$\Delta t_f$ & Fine time-step\\
$\nu$ & Kinematic viscosity\\
$\rho$ & Fluid density\\
$\tau$ & Time (dimensionless) [$\tau\triangleq tU_{\infty}/c$]\\
$\tau_h$ & Hold time\\
$\tau_r$ & Ramp time
\end{tabular}
\end{minipage}}
\end{figure*}

Many aerodynamic models used for flight control rely on the quasi-steady assumption that forces and moments depend in a static manner on parameters such as relative velocity and angle of attack.  In essence, the assumption is that maneuvers are sufficiently slow so that the flow has time to equilibrate.  While these models work well for conventional aircraft, they do not describe the unsteady aerodynamic forces that are important for small, agile aircraft to avoid obstacles, respond to gusts, and track potentially elusive targets.  Small, lightweight aircraft have a lower stall velocity; therefore, gusts and rapid motions excite large reduced frequencies, $k=\pi f c/U_{\infty}$, and Strouhal numbers, $\Str=fM/U_{\infty}$, where $f$ and $M$ are the frequency and amplitude of motion, respectively; lengths are nondimensionalized by the chord length $c$, velocities by the free-stream velocity $U_\infty$ and time by $c/U_\infty$.  Loosely speaking, large reduced frequencies are excited when wing motion is so fast that unsteady flow structures do not have time to convect an entire chord length before new structures are formed.  Large Strouhal numbers are excited by wing motions that are a combination of fast and large amplitude, and these typically result in complex wake structures.  The Strouhal number and reduced frequency may be varied independently by the choice of frequency $f$ and amplitude $M$.

There exist a wide range of unsteady aerodynamic models in the literature~\cite{dowell:2001,leishman:06,Farhat:2010}.  The classical unsteady models of~\cite{wagner:25} and~\cite{theodorsen:35} are still used extensively, and they provide a standard of comparison for the linear models that follow them~\cite{Bruno:2008}.  Wagner's model constructs the lift in response to arbitrary input motion by convolving the time derivative of the motion with the analytically computed step response, also called the \emph{indicial response}.  Linear indicial response models are a mainstay of the unsteady aerodynamics~\cite{tobak:1954} and aeroelasticity~\cite{Marzocca:2002,Costa:2007} communities.  They may be constructed based on analytical, experimental, or numerical step-response information.  They have been applied to a wide range of problems ranging from understanding the effect of control surfaces~\cite{Leishman:1994}, to the modeling of gusts~\cite{Leishman:1996}, and the suppression of shedding on bridges and buildings~\cite{salvatori:2006,Costa:2007}.  Nonlinear extensions have been developed \cite{tobak:85,Prazenica:2007,Balajewicz:2012}.

Theodorsen's frequency-domain model is equivalent to Wagner's, using the same model assumptions of an incompressible, inviscid flow with a planar wake.  Although Theodorsen's model applies only to sinusoidal input motion, it was suitable for the analysis of flutter instability with the tools available.  However, with modern tools, it is possible to construct a state-space realization based on Theodorsen's model that is useful for time-domain analysis and feedback control~\cite{brunton:2012a}.  Accurate state-space aerodynamic models are especially important when the flight dynamic and aerodynamic time scales are comparable.  In this case, modern control techniques such as $\mathcal{H}_{\infty}$-synthesis can be especially useful for achieving robust performance.  Because small, lightweight aircraft have shorter flight dynamic time-scales, small vehicles and bio-flyers at low Reynolds number, $\Rey=cU_{\infty}/\nu$, between $10^2$--$10^5$ are particularly interesting; here $\nu$ is the kinematic viscosity.  However, the classical models are limited by the inviscid assumption that allows for them to be solved in closed form, which makes them less accurate for low Reynolds numbers and at larger angles of attack.

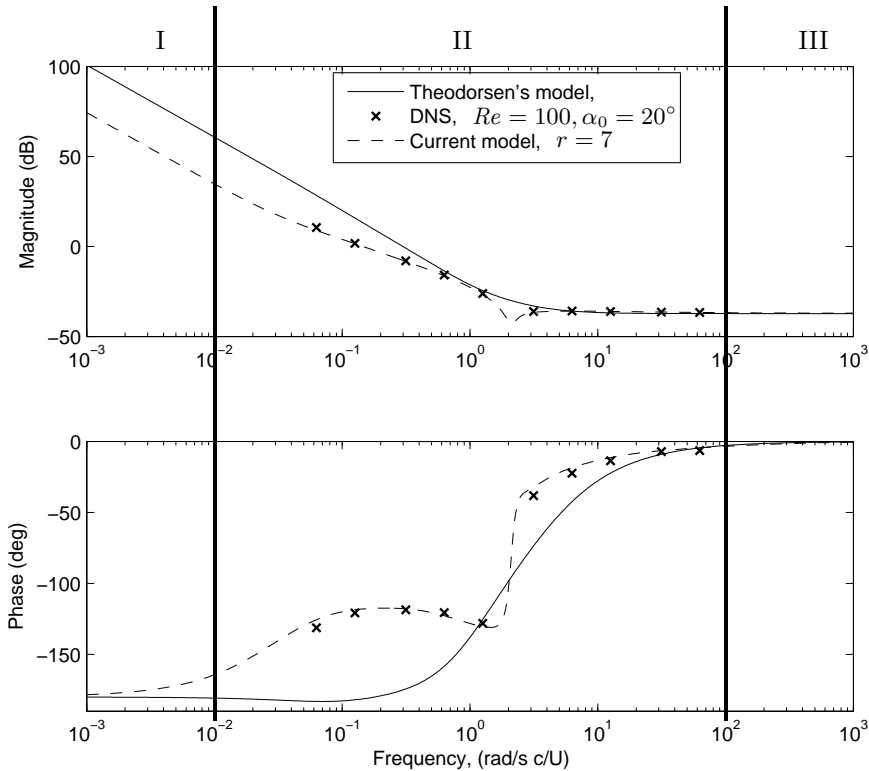
\begin{figure}
\begin{center}
\begin{overpic}[width=.75\textwidth]{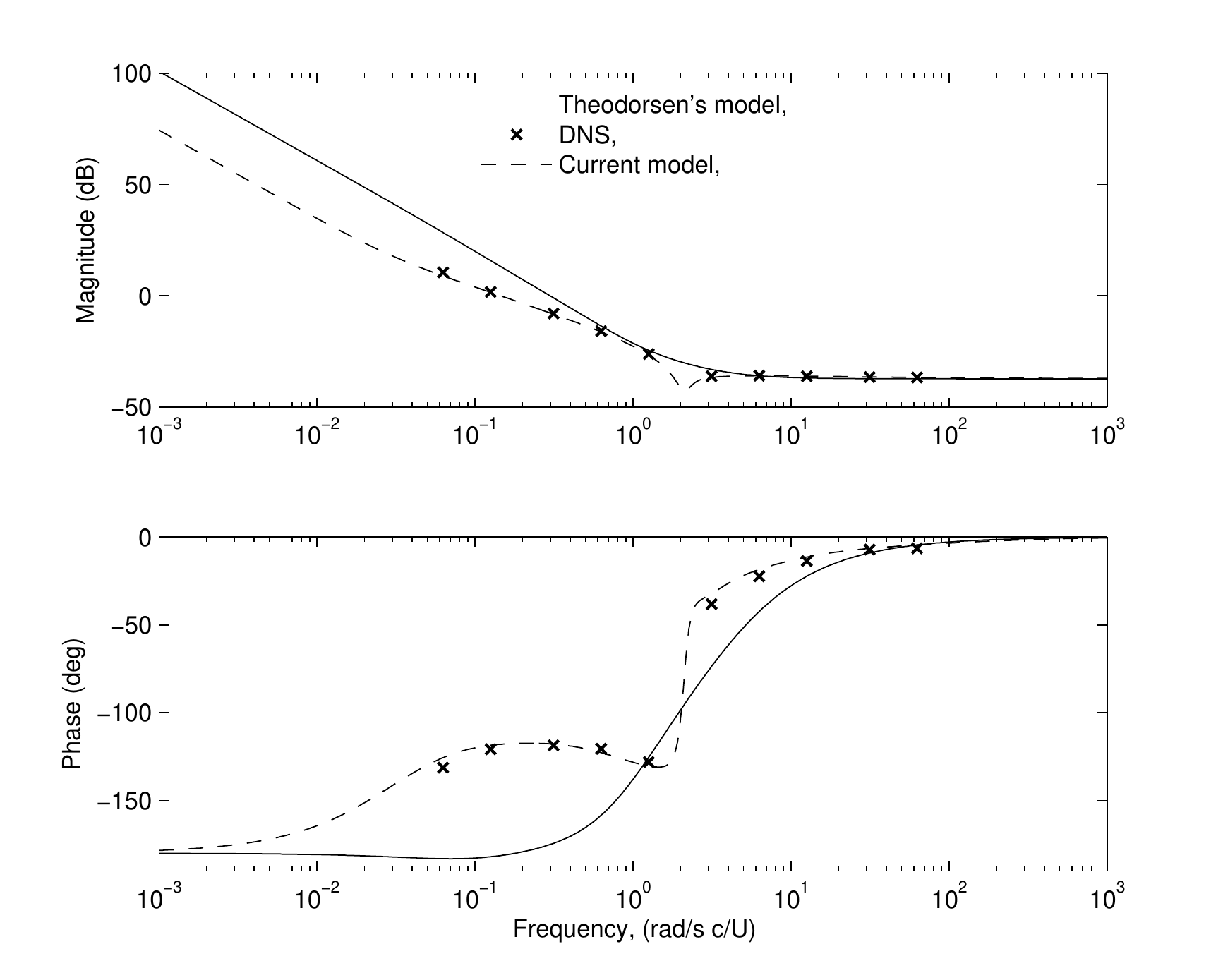}
	\put(25.8,8){\rule{1.2pt}{95mm}} 
	\put(77.5,8){\rule{1.2pt}{95mm}} 
	\put(20,76){\text{I}} 
	\put(50,76){\text{II}} 
	\put(85,76){\text{III}}
	\put(52,68.35){\small$\Rey=100,\alpha_0=20^{\circ}$}
	\put(38,64.5){\framebox(35,9){}}
	\put(60.5,66){\small$r=7$}
\end{overpic}
\vspace{-.22in}
\caption{Bode plot of unsteady lift coefficient, $C_L$, for pitching motion about the leading edge.  The model input is $\ddot\alpha$.  Region I corresponds to low-frequency motion.  Region II corresponds to the transient fluid dynamic frequencies.  Region III corresponds to high-frequency motion dominated by added-mass.  The angle of attack $\alpha$ has units of degrees, and the input motion is $\alpha = \alpha_0 + 1^\circ \sin(2 k t)$.}
\label{fig:bode}
\end{center}
\vspace{-.2in}
\end{figure}

%ERA~\cite{ERA:1985,ERA:2009} yield balanced state space models that are useful for control from impulse response data, which in turn can be obtained from (potentially noisy) input-output data using OKID~\cite{juang:1991}.
Low-dimensional, state-space models for the \emph{viscous} unsteady aerodynamic forces on a small-scale wing in motion were recently developed based on the linearized Navier-Stokes equations with lift-force output~\cite{brunton:2012b}.  
The eigensystem realization algorithm (ERA)~\cite{ERA:1985,ERA:2009} was used to identify balanced reduced-order models from impulse response data, which in turn was obtained from noisy input--output data using the observer/Kalman filter identification (OKID)~\cite{juang:1991}.  
%The eigensystem realization algorithm (ERA) and observer/Kalman filter identification (OKID) were used to identify accurate, efficient reduced-order models from numerical and experimental data.  
The resulting models were inspired by the models of Wagner and Theodorsen, but formulated in state-space and generalized to include transient viscous fluid dynamic forces specific to a particular wing geometry and Reynolds number.  Figure~\ref{fig:bode} shows the frequency response for a flat plate pitching sinusoidally ($\pm 1^\circ$) about its leading-edge at $\Rey=100$ at a base angle of attack of $\alpha_0=20^{\circ}$.  The input is angular acceleration $\ddot\alpha$ and the output is the lift coefficient $C_L=2L/\rho U_{\infty}^2c$, where $L$ is the dimensional lift force and $\rho$ is the fluid density.  Theodorsen's model does not agree with the numerically computed frequency response, except at high frequencies when the forces are dominated by added-mass.  The inviscid model over-predicts the low-frequency magnitude since the model uses a $2\pi$ lift slope.  In addition, the phase is off at intermediate frequencies because Theodorsen's model does not accurately capture the transient viscous dynamics that result from a separation bubble.  
In contrast, a viscous model with a $7$-mode ERA component, based on methods presented in this paper, accurately captures the entire frequency response.

\subsection{Contributions of this work}
This work extends the unsteady aerodynamic models presented in~\cite{brunton:2012b}, investigates how to identify these models from various input--output data, and uses them to develop robust feedback control laws.  Inspired by Theodorsen's model, we introduce a new multiple-input model that is explicitly parameterized by pitch-axis location, and is capable of handling simultaneous pitch, plunge, and surge.  The model order does not increase significantly over a single-input model because of the way that viscous fluid forces are modeled using ERA; the models are shown to be valid up to an effective angle of attack of $\alpha_\text{eff}=20^\circ$, for the example considered.  Next, we investigate the correct uses of ERA/OKID to identify these reduced-order models, and illustrate the potential pitfalls of incorrect modeling.  Finally, robust feedback controllers are developed, based on these low-order state-space models, which are able to attenuate sensor noise and compensate for strong fluid nonlinearities.  

This paper is organized as follows:  In Section~\ref{chap:models:linear}, we summarize previous single-input pitch and plunge models, and then develop a new parameterized multiple-input model.  Section~\ref{chap:models:algorithms} presents an overview of how to identify reduced-order aerodynamic models from input--output data.  New results on multiple-input model identification and the pitfalls of incorrectly applying ERA/OKID are included in Sections~\ref{chap:models:algorithms:MIMO} and~\ref{chap:models:algorithms:naiveokid}.  Section~\ref{sec:results} contains the results of direct numerical simulations of a pitching and plunging flat plate at $\Rey=100$.  The parameterized multi-input model is verified, and robust feedback controllers are developed from linear models and applied to the full non-linear simulations.  The controllers perform well, even when the flow is massively separated.  Section~\ref{chap:models:summary} summarizes the results and future directions.  System identification maneuvers are included in Appendix~\ref{chap:models:maneuvers}.

%%%%%%%%%%%%%%%%%%%%%%%%
%%% MODELS
%%%%%%%%%%%%%%%%%%%%%%%%
\section{State-space unsteady aerodynamic models}\label{chap:models:linear}

The unsteady, incompressible Navier-Stokes equations may be linearized about a nominal angle of attack for various pitch, plunge, and surge motions.  The result is a state-space model that is approximated using system identification and model reduction techniques (see Section~\ref{chap:models:algorithms}).  We also consider a linearized lift-force output equation, although other output measurements are also possible, including forces, moments, pressures, etc.  The models in Sections~\ref{chap:models:linear:pitch} and \ref{chap:models:linear:plunge} have been presented in previous work~\cite{brunton:2012b}.  The new multi-input model in Eq.~(\ref{eq:MIMO2}) is parameterized by pitch-axis location and is inspired by Theodorsen's model.

% FORMULATION I
\subsection{Pitch models}\label{chap:models:linear:pitch}
For the case of pitching motion, linearization results in either of the two models:

\begin{minipage}{0.4\textwidth}
\begin{align}
\frac{d}{dt}\begin{bmatrix} \bsx\\ \alpha\\ \dot\alpha\end{bmatrix} & = \begin{bmatrix} \mA & \m{0} & \mB_{\dot\alpha}\\ \m{0} & 0 & 1\\ \m{0} & 0 & 0\end{bmatrix}\begin{bmatrix} \bsx\\ \alpha\\ \dot\alpha\end{bmatrix} + \begin{bmatrix} \m{0}\\ 0 \\ 1 \end{bmatrix} \ddot\alpha, \nonumber\\
&&\label{eq:rom2}\\
C_L&= \begin{bmatrix} \mC & C_{\alpha} & C_{\dot\alpha}\end{bmatrix} \begin{bmatrix} \bsx\\ \alpha\\ \dot\alpha\end{bmatrix} + C_{\ddot\alpha}\ddot\alpha, \nonumber
\end{align}
\end{minipage}
\begin{minipage}{0.1\textwidth}
~
\end{minipage}
\begin{minipage}{0.4\textwidth}
\begin{align}
\frac{d}{dt}\begin{bmatrix} {{\bsx}}\\ \alpha\\ \dot\alpha\end{bmatrix} & = \begin{bmatrix} \mA & \m{0} & \m{0}\\ \m{0} & 0 & 1\\ \m{0} & 0 & 0\end{bmatrix}\begin{bmatrix} {{\bsx}}\\ \alpha\\ \dot\alpha\end{bmatrix} + \begin{bmatrix} \mB_{\ddot\alpha}\\ 0 \\ 1 \end{bmatrix} \ddot\alpha, \nonumber\\
&&\label{eq:rom3}\\
C_L&= \begin{bmatrix} \mC & C_{\alpha} & C_{\dot\alpha}'\end{bmatrix} \begin{bmatrix} {{\bsx}}\\ \alpha\\ \dot\alpha\end{bmatrix} +  C_{\ddot\alpha}\ddot\alpha. \nonumber
\end{align}
\end{minipage}\\

\noindent The input to the model is $\ddot\alpha$, where $\alpha$ is the angle of attack, the output of the model is the lift coefficient $C_L$, and $\bsx$ is a vector describing the fluid state.
Eqs.~(\ref{eq:rom2}) and (\ref{eq:rom3}) are related by a coordinate transformation so that the state in Eq.~(\ref{eq:rom2}) corresponds to vorticity and the state in Eq.~(\ref{eq:rom3}) corresponds to velocity~\cite{brunton:2012b}.  Notice that in Eq.~(\ref{eq:rom2}) $\dot\alpha$ is the input to the transient $\bsx$ dynamics and in Eq.~(\ref{eq:rom3}) $\ddot\alpha$ is the input to the $\bsx$ dynamics.  This is because the velocity state responds instantaneously to pitch acceleration and vorticity does not.  The coefficient $C_{\alpha}$ is the quasi-steady lift coefficient slope, $C_{\dot\alpha}$ and $C_{\ddot\alpha}$ are added-mass coefficients, and the models $(\mA,\mB_{\dot\alpha},\mC)$ from Eq.~(\ref{eq:rom2}) and $(\mA,\mB_{\ddot\alpha},\mC)$ from Eq.~(\ref{eq:rom3}) are state-space models for the transient fluid dynamic forces.

% PLUNGE MODELS
\subsection{Plunge and surge models}\label{chap:models:linear:plunge}
Models for plunge and surge are related to the pitch models above, except that there is no steady-state force associated with a specific horizontal position $g$ or vertical position $h$;  thus, there are no $C_g$ or $C_h$ coefficients in the model.  Because $\ddot{g}$ and $\ddot{h}$ contribute to the rate of effective angle of attack, $\dot\alpha_{e}$, these generate vorticity and are considered to be the inputs to the transient $\bsx$-dynamics.  A simple plunge model is given by:

\begin{align}
\frac{d}{dt}\begin{bmatrix}\bsx\\ \dot{h}\end{bmatrix}&= \begin{bmatrix}\mA & \m{0}  \\ \m{0} & 0\end{bmatrix}\begin{bmatrix}\bsx\\ \dot{h} \end{bmatrix} + \begin{bmatrix} \mB_{\ddot h}  \\ 1\end{bmatrix} \ddot{h}\nonumber\\
&\label{eq:plunge}\\
C_L&= \begin{bmatrix} \mC & C_{\dot{h}} \end{bmatrix}\begin{bmatrix}\bsx\\ \dot{h} \end{bmatrix}  + C_{\ddot{h}}~ \ddot{h}. \nonumber
\end{align}
 Surge is identical, with $g$ replacing $h$.

\subsection{Combined pitch, plunge and surge models}\label{chap:models:linear:combined}
It is possible to combine the pitch, plunge and surge models above into a single state-space representation.  We may use either pitch formulation above (with either $\dot\alpha$ or $\ddot\alpha$ as input to the state $\bsx$).  Combining the pitch model in Eq.~(\ref{eq:rom3}) with plunge and surge models in Eq.~(\ref{eq:plunge}) yields the following multiple-input, single-output (MISO) model:
\begin{align}
\displaystyle{\frac{d}{dt}}\begin{bmatrix} \bsx\\ \alpha \\ \dot\alpha\\ \dot h\\ \dot g\end{bmatrix}
& = \begin{bmatrix} 
	\mA & \m{0} & \m{0} & \m{0} & \m{0} \\ 
	\m{0} & 0 & 1 & 0 & 0\\
	\m{0} & 0 & 0 & 0 & 0\\
	\m{0} & 0 & 0 & 0 & 0\\
	\m{0} & 0 & 0 & 0 & 0
	\end{bmatrix}
	\begin{bmatrix}\bsx \\ \alpha \\ \dot\alpha\\ \dot h\\ \dot g\end{bmatrix}
	+ \begin{bmatrix}
	\mB_{\ddot\alpha} & \mB_{\ddot h} & \mB_{\ddot g}\\
	0 & 0 & 0\\
	1 & 0 & 0\\
	0 & 1 & 0\\
	0 & 0 & 1
	\end{bmatrix}\begin{bmatrix} \ddot\alpha \\ \ddot h\\ \ddot g\end{bmatrix}\nonumber\\
&\label{eq:MIMO}\\
C_L&=\begin{bmatrix} \mC & C_{\alpha} & C_{\dot\alpha} & C_{\dot h} & C_{\dot g}\end{bmatrix}
\begin{bmatrix} \bsx \\ \alpha \\ \dot\alpha \\ \dot h \\ \dot g\end{bmatrix} 
+ \begin{bmatrix} C_{\ddot\alpha} & C_{\ddot h} & C_{\ddot g}\end{bmatrix}\begin{bmatrix} \ddot\alpha \\ \ddot h\\ \ddot g\end{bmatrix}.\nonumber
\end{align}

The model in Eq.~(\ref{eq:MIMO}) is linearized about a base angle of attack $\alpha_0$ for a \textit{specific} pitch-axis location $a$, which is measured with respect to the $1/2$-chord (e.g., pitching about the leading edge corresponds to $a=-1/2$, whereas the trailing edge is $a=1/2$).  However, it is possible to obtain a model that is parameterized by the pitch-axis location $a$.  All pitch motions about a given point $a$ may be considered a combination of pitch about the mid-chord (or any point of interest) and an induced plunge and surge motion.  The magnitudes of the induced plunge and surge motions at the middle-chord location are $a\mathcal{C}_0\ddot\alpha$ and $a\mathcal{S}_0\ddot\alpha$, respectively, where $\mathcal{C}_0=\cos(\alpha_0)$, and $\mathcal{S}_0=\sin(\alpha_0)$.  Note that this is not an approximation, but is an exact kinematic transformation; this idea originates from~\cite{theodorsen:35}, but is generalized here to be valid for large angles of attack.  A model parameterized by the pitch-axis location~$a$ is given by the following MISO model:  
\begin{align}
\frac{d}{dt}\begin{bmatrix} \bsx\\ \alpha \\ \dot\alpha\\ \dot h\\ \dot g\end{bmatrix}
& = \begin{bmatrix} 
	\mA & \m{0} & \m{0} & \m{0} & \m{0} \\ 
	\m{0} & 0 & 1 & 0 & 0\\
	\m{0} & 0 & 0 & 0 & 0\\
	\m{0} & 0 & 0 & 0 & 0\\
	\m{0} & 0 & 0 & 0 & 0
	\end{bmatrix}
	\begin{bmatrix}\bsx \\ \alpha \\ \dot\alpha\\ \dot h\\ \dot g\end{bmatrix}
	+ \begin{bmatrix}
	\mB_{\ddot\alpha}+a(\mathcal{C}_{0}\mB_{\ddot h} +\mathcal{S}_{0}\mB_{\ddot g}) & \mB_{\ddot h} & \mB_{\ddot g}\\
	0 & 0 & 0\\
	1 & 0 & 0\\
	a\mathcal{C}_{0} & 1 & 0\\
	a\mathcal{S}_{0} & 0 & 1
	\end{bmatrix}\begin{bmatrix} \ddot\alpha \\ \ddot h\\ \ddot g\end{bmatrix}\nonumber\\
&\label{eq:MIMO2}\\
C_L&=\begin{bmatrix} \mC & C_{\alpha} & C_{\dot\alpha} & C_{\dot h} & C_{\dot g}\end{bmatrix}
\begin{bmatrix} \bsx \\ \alpha \\ \dot\alpha \\ \dot h \\ \dot g\end{bmatrix} 
+ \begin{bmatrix} 
C_{\ddot\alpha} + a(\mathcal{C}_{0}C_{\ddot h} +\mathcal{S}_{0}C_{\ddot g})& C_{\ddot h} & C_{\ddot g}\end{bmatrix}\begin{bmatrix} \ddot\alpha \\ \ddot h\\ \ddot g\end{bmatrix}.\nonumber
\end{align}
Here, $(\mA,\mB_{\ddot\alpha} ,\mB_{\ddot h}, \mB_{\ddot g}, \mC)$ are from the model in Eq.~(\ref{eq:MIMO}) for the case of pitch about the mid-chord.  Mid-chord pitching is particularly simple for a symmetric plate, since there are no added-mass lift forces from $\ddot\alpha$, so $C_{\ddot\alpha}=0$.  

Alternatively, one may use body-frame coordinates $(\xi,\eta,\alpha)$, where $\xi$ is parallel to the airfoil, $\eta$ is normal, and $\alpha$ is the pitch angle.  At a base angle $\alpha_0$, the induced acceleration at the mid-chord is entirely in the direction $\eta$:
%This results in a slightly simpler model:
\begin{align}
\frac{d}{dt}\begin{bmatrix} \bsx\\ \alpha \\ \dot\alpha\\ \dot \eta\\ \dot\xi\end{bmatrix}
& = \begin{bmatrix} 
	\mA & \m{0} & \m{0} & \m{0} & \m{0} \\ 
	\m{0} & 0 & 1 & 0 & 0\\
	\m{0} & 0 & 0 & 0 & 0\\
	\m{0} & 0 & 0 & 0 & 0\\
	\m{0} & 0 & 0 & 0 & 0
	\end{bmatrix}
	\begin{bmatrix}\bsx \\ \alpha \\ \dot\alpha\\ \dot \eta\\ \dot \xi\end{bmatrix}
	+ \begin{bmatrix}
	\mB_{\ddot\alpha}+a\mB_{\ddot\eta} & \mB_{\ddot\eta} & \mB_{\ddot\xi}\\
	0 & 0 & 0\\
	1 & 0 & 0\\
	a & 1 & 0\\
	0& 0 & 1
	\end{bmatrix}\begin{bmatrix} \ddot\alpha \\ \ddot \eta\\ \ddot \xi\end{bmatrix}\nonumber\\
&\label{eq:MIMO3}\\
C_L&=\begin{bmatrix} \mC & C_{\alpha} & C_{\dot\alpha} & C_{\dot \eta} & C_{\dot \xi}\end{bmatrix}
\begin{bmatrix} \bsx \\ \alpha \\ \dot\alpha \\ \dot \eta \\ \dot \xi\end{bmatrix} 
+ \begin{bmatrix} 
C_{\ddot\alpha} + aC_{\ddot \eta} & C_{\ddot \eta} & C_{\ddot \xi}\end{bmatrix}\begin{bmatrix} \ddot\alpha \\ \ddot \eta\\ \ddot \xi\end{bmatrix}.\nonumber
\end{align}
The $\mB$ and $\mD$ matrices in Eq.~(\ref{eq:MIMO3}) are simple compared with those in Eq.~(\ref{eq:MIMO2}); in the case of $\alpha_0=0^{\circ}$, they are equal.
The $\xi$ term may often be neglected for a flat plate, since parallel motion generates significantly less induced fluid velocity than normal motion, meaning that $\mB_{\ddot\xi}$, $C_{\dot\xi}$ and $C_{\ddot\xi}$ are generally small compared with $\mB_{\ddot\eta}$, $C_{\dot\eta}$ and $C_{\ddot\eta}$.  
%  Often $\mB_{\ddot\xi}$ is much smaller than $\mB_{\ddot\eta}$ since parallel motion generates significantly less induced velocity from the plate than normal motion.  Thus, the $\xi$ term may often be neglected, further simplifying Eq.~(\ref{eq:MIMO3}).

Note that the various $\mA$ and $\mC$ matrices above are not necessarily equal to one another, although they are equal for Eqs.~(\ref{eq:MIMO}) and (\ref{eq:MIMO2}).  The $\mA,\mB_i$, and $\mC$ matrices and the coefficients $C_{\dot\alpha}$ and $C_{\ddot\alpha}$ in Eqs.~(\ref{eq:rom2}), (\ref{eq:rom3}), and (\ref{eq:MIMO}) depend on the pitch-axis location $a$.  Finally, the matrices and constants for every model in this paper depend on the base angle of attack $\alpha_0$, wing geometry, and Reynolds number.  A remarkable property is that the multiple-input models tend not to require increased model order of the $\bsx$-dynamics, indicating that the reduced-order models for pitch and plunge motions share similar transient viscous fluid dynamics, at least for the examples considered here.

%%%%%%%%%%%%%%%%%%%%%%%%
%%% ALGORITHMS 
%%%%%%%%%%%%%%%%%%%%%%%%
\section{Reduced-order aerodynamic modeling procedure}\label{chap:models:algorithms}
The algorithms in this section provide a methodology for obtaining low-dimensional representations for the unsteady aerodynamic models in Section~\ref{chap:models:linear} from either numerical or experimental data.  There are three basic algorithms which may be extended or modified as necessary.  These methods are presented for a pitching input ($u=\ddot\alpha$) and lift coefficient output ($y=C_L$).  However, they may be generalized to include plunge and surge input motions, as well as drag and moment coefficient outputs, as discussed in Section~\ref{chap:models:algorithms:MIMO}.  These algorithms have been used previously to obtain single-input models~\cite{brunton:2012b}, although the details presented here are new.  Sections~\ref{chap:models:algorithms:MIMO} and~\ref{chap:models:algorithms:naiveokid} present new material on multiple-input models and modeling pitfalls, respectively.

The following methods rely heavily on the eigensystem realization algorithm (ERA) and the observer/Kalman filter identification (OKID) method.  The ERA was developed by~\cite{ERA:1985} to extend the minimal realization theory of~\cite{kalman:1965} to systems with noisy data.  The ERA produces reduced-order models of a linear time-invariant system based on its Markov parameters, $\mathcal{H}_i$, which are the output history from impulse-response experiments of a discrete-time system.  When these parameters decay slowly, as in lightly damped systems, the ERA involves computing the singular value decomposition of a large  matrix.  For this reason, the ERA is often used in conjunction with OKID~\cite{juang:1991}, which constructs an asymptotically stable observer to identify the system.  OKID identifies the system Markov parameters using input--output data from realistic maneuvers, such as actual flight data, as in \cite{valasek:2003}.  

Recently, \cite{ERA:2009} have shown that the ERA efficiently produces the same reduced-order models as the method of snapshot-based balanced proper orthogonal decomposition (BPOD)~\cite{rowley:05pod}, without the need for adjoint simulations.  Balanced models have been effectively used for feedback control on a number of physical applications including the flat-plate boundary layer~\cite{bagheri:2009}, a flat plate at high incidence~\cite{ahuja:2010}, cavity resonances and combustion oscillations~\cite{illingworth:2010}, and the transitional channel flow~\cite{ilak:2008}.  The work of \cite{Silva:2004} has used the ERA and OKID to obtain reduced-order models from CFD for the unsteady aerodynamics excited by aeroelastic modes.  In many examples, the performance of balanced models is striking when compared against Galerkin projection onto POD modes.

It may appear at first glance that it would be simplest to use the ERA/OKID method to identify the entire model in Eq.~(\ref{eq:rom2}) or Eq.~(\ref{eq:rom3}).  However, the structure of the model, with added-mass forces proportional to $\ddot\alpha$ and quasi-steady forces proportional to $\alpha$, makes it important to subtract these off before modeling the remaining \textit{transient dynamics} with the ERA.  This concept is illustrated in Section~\ref{chap:models:algorithms:naiveokid}.

\subsection{Common features of all methods}\label{chap:models:algorithms:general}
All of the algorithms below share some common features.  Each is based on a discrete-time impulse response in either $\dot\alpha$ or $\ddot\alpha$, which corresponds to a step response in either $\alpha$ or $\dot\alpha$, respectively.  The step response may be obtained directly or estimated from a frequency-rich input--output maneuver using the OKID method.  Various maneuvers for system identification are presented in Section~\ref{chap:models:maneuvers}.  

\begin{figure}
\begin{center}
\begin{tabular}{ccc}
\begin{overpic}[height=.45\textwidth]{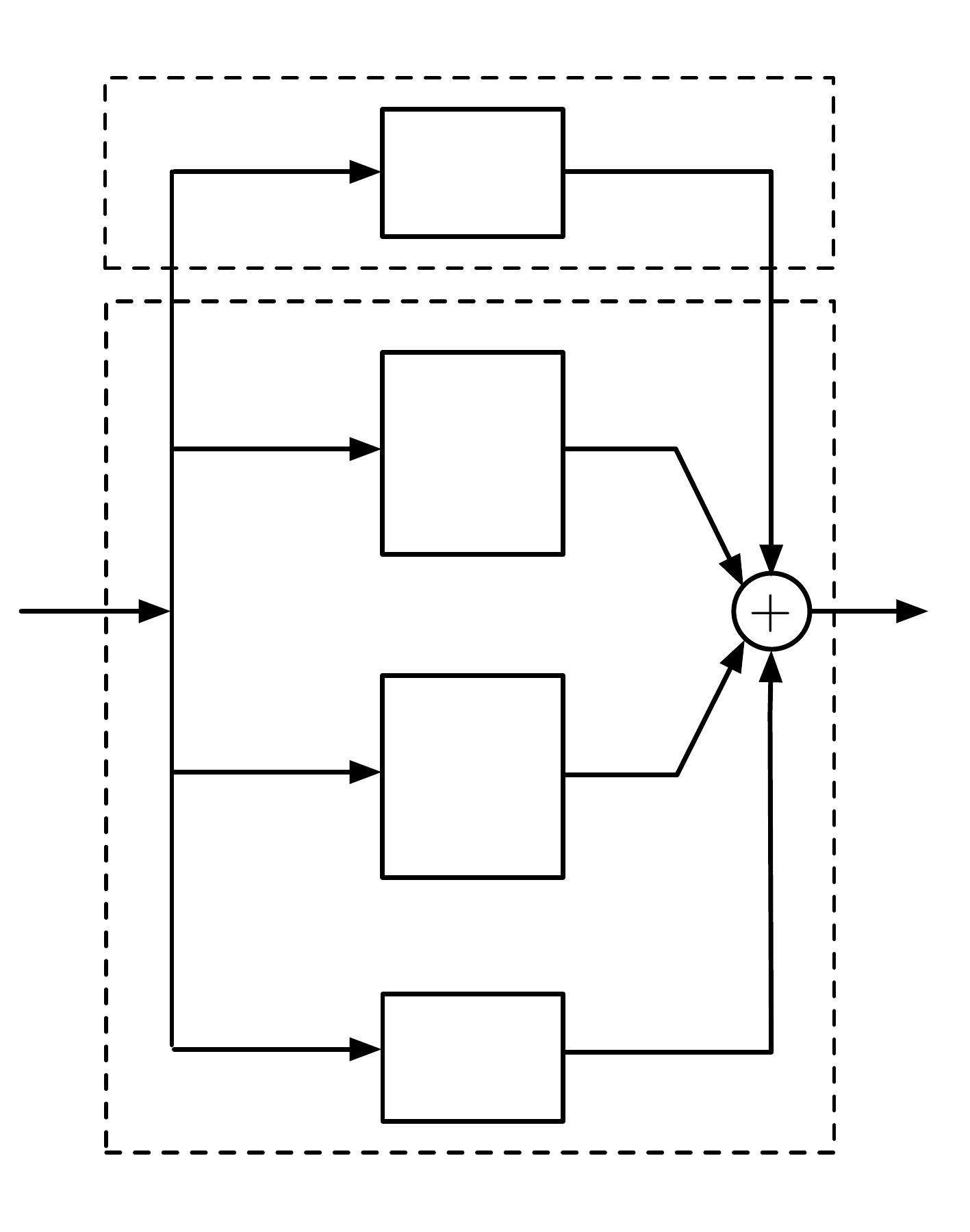}
\put(35,13){{$C_{\ddot\alpha}$}}
\put(34.5,35){{$\dfrac{C_{\dot\alpha}}{s}$}}
\put(34.5,61.5){{$\dfrac{C_{\alpha}}{s^2}$}}
\put(32.6,84.5){{$\dfrac{1}{s}G(s)$}}
\put(2,52.5){{$\ddot\alpha$}}
\put(68.5,53){{$C_L$}}
\put(10,2){{\small quasi-steady \& added-mass}}
\put(27,95){{\small transient dynamics}}
\put(2,2){(a)}
\end{overpic}
&~~~~&
\begin{overpic}[height=.45\textwidth]{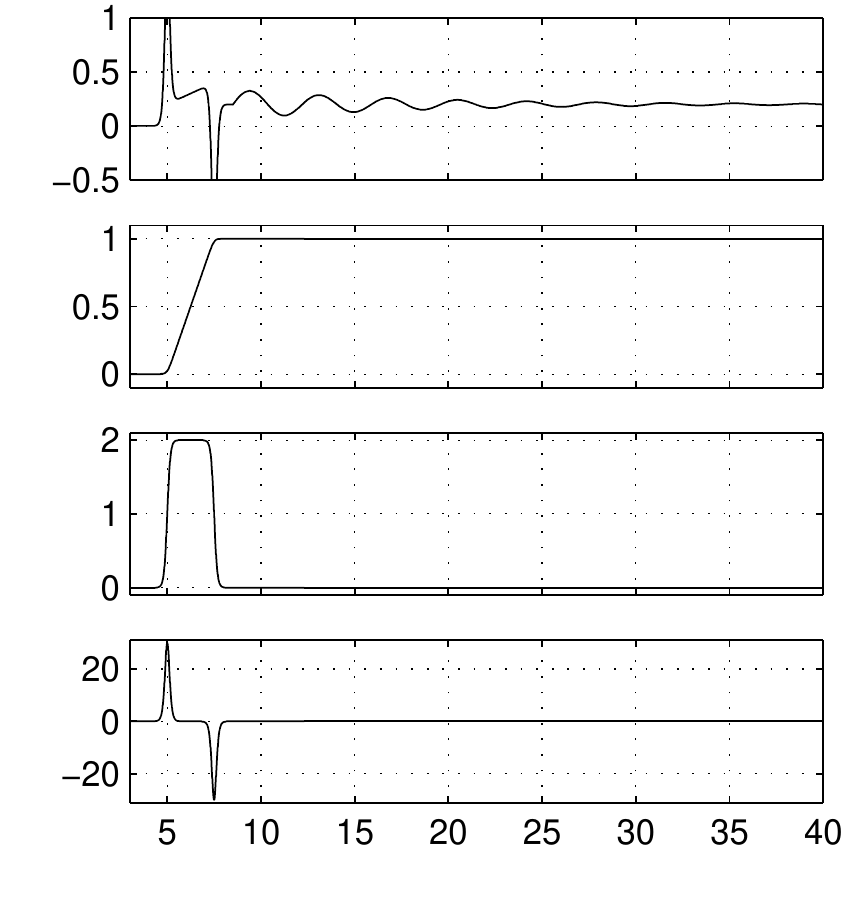}
\put(15,94){\bf{1}}
\put(20,91){\bf 2}
\put(35,90){\bf 3}
\put(70,90){\bf 4}
\put(25,1){{\small Convective time, ($\tau=tU/c$)}}
\put(0,19.5){\begin{sideways}{$\ddot\alpha$}\end{sideways}}
\put(0,42){\begin{sideways}{$\dot\alpha$ }\end{sideways}}
\put(0,59.5){\begin{sideways}{$\alpha$ (deg)}\end{sideways}}
\put(0,85){\begin{sideways}{$C_L$}\end{sideways}}
\put(85,92.5){$\bs{Y}$}
\put(85,24){$\bs{U}$}
\put(2,2){(b)}
\end{overpic}
\end{tabular}
\end{center}
\vskip -.1in
\caption{(a) Schematic for reduced-order model (\ref{eq:rom2}), where $G(s)=\mC(s\m{I}-\mA)^{-1}\mB_{\dot\alpha}$. (b) Sketch of a typical aerodynamic step response for pitch about the leading edge.  (top) Lift coefficient decomposed into: 1. added-mass proportional to $\ddot\alpha$, 2. added-mass proportional to $\dot\alpha$ plus quasi-steady forces (proportional to $\alpha$), 3. transient dynamics plus quasi-steady forces, and 4. quasi-steady forces.}\label{fig:model:schematic}
\end{figure}
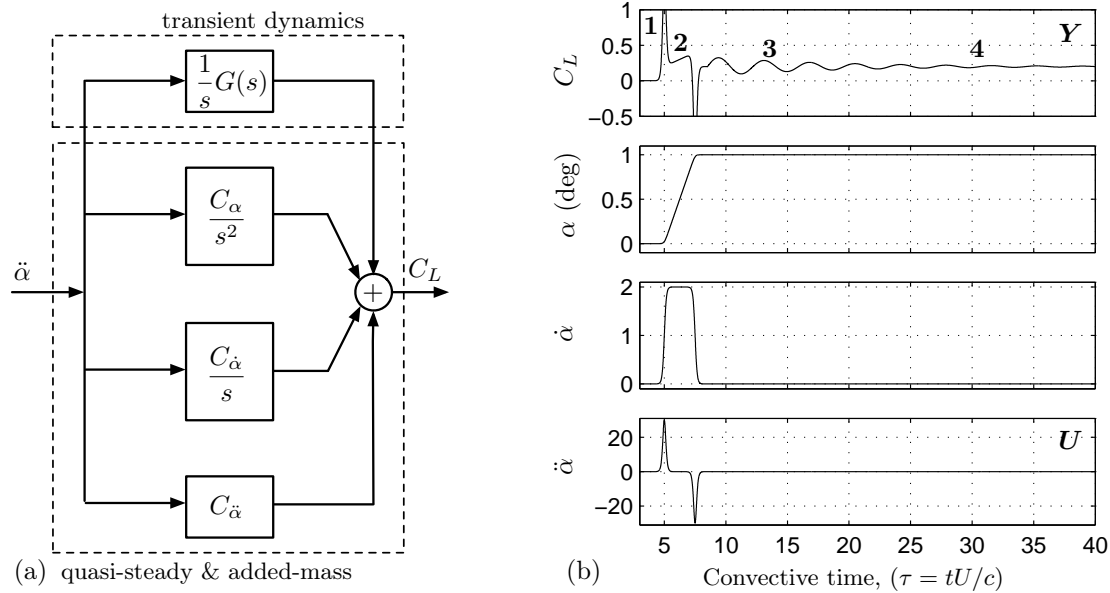

After obtaining a discrete-time step response in either $\alpha$ or $\dot\alpha$, the quasi-steady and added-mass coefficients $C_{\alpha}, C_{\dot\alpha}$, and $C_{\ddot\alpha}$ are identified and subtracted from the measured response.  After subtracting off these effects, the last step is to identify the remaining transient dynamics using the ERA.  The transient portion $(\mA,\mB,\mC)$ of the model in Eqs.~(\ref{eq:rom2}) and~(\ref{eq:rom3}) is approximated by the ERA model $(\mA,\mB,\mC)_r$ of order $r\ll n=\text{dim}(\mA)$; the input is either $\dot\alpha$ or $\ddot\alpha$, and the output is $C_L$.  This splitting of quasi-steady, added-mass, and transient effects is illustrated in Figure~\ref{fig:model:schematic} (a) for Eq.~(\ref{eq:rom2}), where $G(s)=\mC(s\m{I}-\mA)^{-1}\mB_{\dot\alpha}$ is the transfer function for the transient dynamics.  The general procedure is summarized as follows:
\begin{enumerate}
\item Obtain the impulse response in either $\dot\alpha$ or $\ddot\alpha$, possibly via OKID.
\item Determine $ C_{\alpha}$, $C_{\dot\alpha}$, and $C_{\ddot\alpha}$.  The coefficients $C_{\alpha}$ and $C_{\ddot\alpha}$ guarantee correct low-frequency and high-frequency behavior of the model, respectively.
\item Identify a reduced-order model for remaining dynamics with the ERA.
\end{enumerate}

Identification of $C_{\alpha}, C_{\dot\alpha}$ and $C_{\ddot\alpha}$ is discussed in Section~\ref{chap:models:algorithms:1}.  $C_{\dot\alpha}$ is determined explicitly in method 1, but this is optional in the other algorithms.  When $\ddot\alpha$ is the input to the transient dynamics, $(\mA,\mB_{\ddot\alpha},\mC)$, as in Eq.~(\ref{eq:rom3}), the $C_{\dot\alpha}$ contribution is captured by $\lim_{s\rightarrow 0}\mC(s\m{I}-\mA)^{-1}\mB_{\ddot\alpha}=-\mC\mA^{-1}\mB_{\ddot\alpha}$.

Before providing details of each specific method, it is helpful to consider a typical step response for an unsteady aerodynamic system modeled by Eq.~(\ref{eq:rom2}) or Eq.~(\ref{eq:rom3}).  The actual aerodynamic step response has many temporal and spatial scales, due to the order of magnitude difference in various forces at different frequencies.  Therefore, to make the individual features more apparent, we use a slower version of the step maneuver from Section~\ref{chap:models:maneuvers:step} on a modified system with many of the same features as the aerodynamic systems we seek to model.  

Figure~\ref{fig:model:schematic} (b) shows a typical ramped step response in $\alpha$ for a wing pitching about a point ahead of the mid-chord.  The top plot shows the lift coefficient history throughout the step, and the bottom three plots show the angle of attack and its derivatives throughout the maneuver.  The step response is characterized by large added-mass forces during the step (1 and 2), followed by a transient lift (3) which decays to a steady-state value (4) after a large number of convective time units.  The added-mass forces are a combination of terms proportional to $\dot\alpha$ and $\ddot\alpha$, and may be written as $ C_{\dot\alpha}  \dot\alpha+C_{\ddot\alpha}  \ddot\alpha$.  The steady-state lift is given by $ C_{\alpha}  \alpha$, where $ C_{\alpha}$ is the lift coefficient slope.  Finally, the transient lift in region $(3)$ comes from unsteady fluid dynamic effects, for example due to separation bubble dynamics, or other boundary layer effects.  This may be represented by $\mC \bsx$, where $\bsx$ represents the generalized fluid dynamic state.  This yields the following:
\begin{eqnarray}
C_L(\alpha,\dot\alpha,\ddot\alpha,\bsx)= C_{\alpha}\alpha
+  C_{\dot\alpha}\dot\alpha +
 C_{\ddot\alpha}\ddot\alpha + \mC\bsx.
\label{eq:stabderiv}
\end{eqnarray}
The coefficients $C_{\alpha,\dot\alpha,\ddot\alpha}$ in Eq.~(\ref{eq:stabderiv}) are related to the stability derivates $C_{L_{\alpha}}\triangleq \partial C_L/\partial \alpha$, $C_{L_{\dot\alpha}}\triangleq \partial C_L/\partial \dot\alpha$, and $C_{L_{\ddot\alpha}}\triangleq \partial C_L/\partial \ddot\alpha$ and the expression is generalized to include additional dynamics via the state $\bsx$.

In the following, $\Delta t_f$ is the time step for time-resolved measurements and $\Delta t_c$ is the time step for coarse sampling and the resulting discrete-time model.  Further, let $\bs{Y}=\begin{bmatrix}  Y_0 & Y_1 & \dotsc & Y_N\end{bmatrix}^T$ be a vector of lift coefficient outputs, $Y_k=C_L(k \Delta t_f)$, and $\bs{U}=\begin{bmatrix} U_0 & U_1 & \dotsc & U_N\end{bmatrix}^T$ be a vector of inputs, $U_k=\ddot\alpha(k\Delta t_f)$, measured at times $k\Delta t_f$.  The number of samples, $N$, must be sufficiently large that transients have died out.

% ALGORITHM 1
\subsection{Method 1: Model (\ref{eq:rom2}) from ramped-step response in $\alpha$}\label{chap:models:algorithms:1}
This method is based on the lift output in response to a ramped step maneuver in $\alpha$ from $\alpha_0$ to $\alpha_0+M$, and is used primarily with direct numerical simulations to identify models of the form in Eq.~(\ref{eq:rom2}).  The step maneuver described in Section~\ref{chap:models:maneuvers:step} is a smoothed linear ramp function that may be viewed as a time-resolved, smoothed approximation to a discrete-time impulse in $\dot\alpha$ of magnitude $M/\Delta t_c$, where $\Delta t_c$ is the duration of the step maneuver.  A small step amplitude, $M\in[0.1^{\circ},1^{\circ}]$, and duration, $\Delta t_c\in[0.01,0.1]$ convection time units, yields a response that is approximately linear and has sufficiently large high-frequency transients.

The lift slope $C_{\alpha}(\alpha_0)$ is equal to $(Y_N-Y_0)/M$, where $Y_N$ is the steady-state lift, measured long after the step, and $Y_0$ is the initial lift at a fixed angle $\alpha_0$ before the step.  We subtract $C_{\alpha}\alpha(k\Delta t_f)$ from each $Y_k$.  

The added-mass coefficient $C_{\ddot\alpha}$ may be solved for in the equation $\bs{Y}\approx C_{\ddot\alpha}\bs{U}$, which is  approximately true during the step maneuver when $\bs{U}$ is large.  We find a least squares fit for $\bs{Y}$ in terms of $\bs{U}$ by taking the pseudoinverse: $C_{\ddot\alpha}= \bs{U}^{\dagger} \bs{Y}\triangleq \bs{U}^*(\bs{U} \bs{U}^*)^{-1} \bs{Y}$.  It is important to use only portions of $\bs{Y}$ and $\bs{U}$ restricted to the step maneuver where the added-mass forces dominate.  We now subtract $C_{\ddot\alpha}U_k$ from each $Y_k$.  

After subtracting off $ C_{\alpha}\alpha$ and $C_{\ddot\alpha}\ddot\alpha$ from the step response, the remaining transient dynamics may be modeled using the ERA.  In order to determine a model using the ERA, we require Markov parameters $\mathcal{H}_i$, which are the outputs $y_k$ from a discrete-time impulse response.  It is possible to obtain $\mathcal{H}_i$ by sampling the signal with time step $\Delta t_c$ starting from the middle of the discrete impulse in $\dot\alpha$.  With the Markov parameters, it is possible to identify the remaining portion of the model corresponding to $C_{\dot\alpha}\dot\alpha + \mC \bsx$:
\begin{itemize}
\item $C_{\dot\alpha}=\mathcal{H}_0\Delta t_c/M$,  (since the magnitude of the discrete pulse in $\dot\alpha$ is $M/\Delta t_c$),
\item $\{\mathcal{H}_{j}\Delta t_c/M\,|\,j\geq 1\}~\rightarrow~ (\mA,\mB_{\dot\alpha},\mC)_r$, from Eq.~(\ref{eq:rom2}), using the ERA.
\end{itemize}
Finally, the reduced-order, discrete-time model $(\mA,\mB_{\dot\alpha},\mC)_r$ may be converted to continuous time, since the dynamics at very high frequencies are dominated by added-mass terms.

As an aside, it is possible to refine the estimate for $C_{\ddot\alpha}$ by subtracting off $C_{\dot\alpha}\dot\alpha$ and correcting for the change in angle of attack during the step: $C_{\ddot\alpha}=\bs{U}^\dagger\tilde{\bs{Y}}$ where $\tilde{\bs{Y}}=\begin{bmatrix} \tilde{Y}_0 & \tilde{Y_1} & \dotsc  & \tilde{Y}_S\end{bmatrix}^T$ and $\tilde Y_k=(Y_k-C_{\dot\alpha}\dot\alpha(k\Delta t_f))/\cos(\alpha(k\Delta t_f))$.  However, the effects of $C_{\alpha}\alpha$ and $C_{\dot\alpha}\dot\alpha$ are generally orders of magnitude smaller than the added-mass force $C_{\ddot\alpha}\ddot\alpha$ during the step, and may be neglected in practice.  

% ALGORITHM 2
\subsection{Method 2: Model (\ref{eq:rom3}) from ramped-step response in $\alpha$}\label{chap:models:algorithms:2}
To obtain a model of the form in Eq.~(\ref{eq:rom3}), it is not practical to directly obtain an impulse in $\ddot\alpha$, since this corresponds to a step in $\dot\alpha$ and a ramp in $\alpha$; constant growth of $\alpha$ will quickly take us out of the linear regime.  Instead, we start with a ramped-step in $\alpha$ and subtract off the quasi-steady lift coefficient, $C_{\alpha}$, as in the previous section.  Next, the coefficient $C_{\dot\alpha}$ may be identified by sampling the measured lift during the middle of the impulse in $\dot\alpha$, when $\ddot\alpha$ is zero. To identify $C_{\ddot\alpha}$, the remaining signal is integrated to give the step response in $\dot\alpha$ (impulse response in $\ddot\alpha$), less the $C_{\alpha}$ and $C_{\dot\alpha}$ contributions.  Sampling the remaining signal yields the Markov parameters $\mathcal{H}_i$ for an impulse in $\ddot\alpha$, which are synthesized into $C_{\ddot\alpha}$ and a low-order, discrete-time model $(\mA,\mB_{\ddot\alpha},\mC)_r$ for the transient dynamics $(\mA,\mB_{\ddot\alpha},\mC)$ in Eq.~(\ref{eq:rom3}).

The signal might have a steady-state value after integration, which may be removed and added to the coefficient $C_{\dot\alpha}$; if it is not removed manually, then it will be absorbed in the term $-\mC\mA^{-1}\mB_{\ddot\alpha}$, using the matrices in Eq.~(\ref{eq:rom2}).  Similarly, it is not strictly necessary to identify and remove the $C_{\dot\alpha}$ term before integrating, as this will also be captured by the transient dynamics.  This is simply a statement of the fact that state-space representations are not unique.  Alternatively, we may consider the Laplace transforms of the lift coefficient, $Y(s)=\mathcal{L}[C_L(t)]$, and angular acceleration, $U(s)=\mathcal{L}[\ddot\alpha(t)]$, resulting in the unique transfer function for Eq.~(\ref{eq:rom3}):
\begin{eqnarray}
Y(s)=\left[\frac{ C_{\alpha}}{s^2} + \frac{ C_{\dot\alpha}}{s}
  +  C_{\ddot\alpha} + G(s) \right] U(s),
\label{eq:IRtf}
\end{eqnarray}
where $G(s)=\mC(s\m{I}-\mA)^{-1}\mB_{\ddot\alpha}$ is a stable, strictly proper transfer function for the additional transient dynamics.  

% ALGORITHM 3
\subsection{Method 3: Model (\ref{eq:rom3}) from impulse response in $\ddot\alpha$ (OKID)}\label{chap:models:algorithms:3}
This method develops models of the form in Eq.~(\ref{eq:rom3}) from realistic input--output maneuvers, such as those developed in Section~\ref{chap:models:maneuvers:okid}.  In particular, the OKID method is used to obtain the Markov parameters for the \emph{linearized} impulse response in $\ddot\alpha$.  From the impulse response, one may identify the parameters $C_{\alpha}, C_{\dot\alpha}$, and $C_{\ddot\alpha}$ as well as a low-dimension ERA model $(\mA,\mB_{\ddot\alpha},\mC)_r$ using a technique similar to the method in Section~\ref{chap:models:algorithms:2}.    

The OKID method provides Markov parameters for a discrete-time impulse response in $\ddot\alpha$ given the input--output data for a frequency-rich maneuver, such as the maneuvers in Section~\ref{chap:models:maneuvers:okid}.  Typically the input--output pair used with OKID have been sampled from the time-resolve input motion $\ddot\alpha(t)$ and lift signal $C_L(t)$.  The sample time $\Delta t_c$ is the desired coarse time step for the discrete-time model of the transient dynamics.  For wind-tunnel experiments, a time step of $\Delta t_c=0.1$ convective time is sufficiently fast to correctly identify added-mass forces.  It has been determined that this time step is small enough to capture the relevant frequencies for the transient dynamics, since higher frequency motions are dominated by added-mass forces, which are accurately captured by the coefficient $C_{\ddot\alpha}$ in the continuous-time model, Eq.~(\ref{eq:rom3}).  

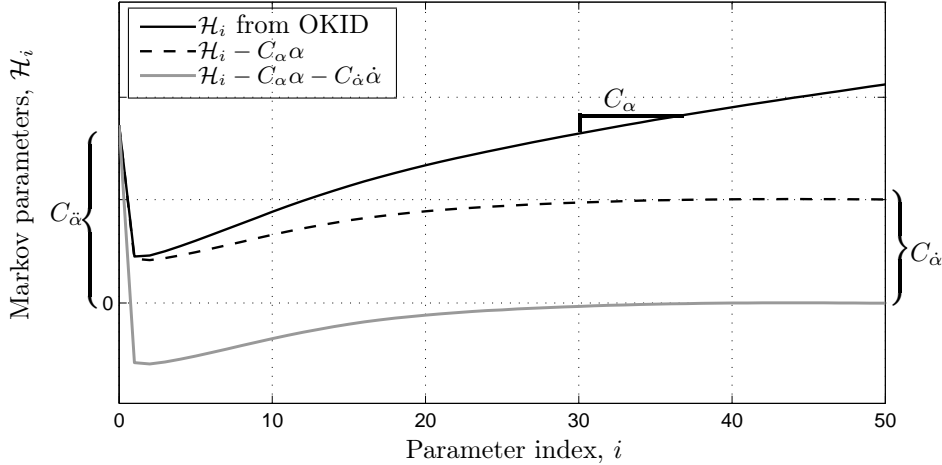
\begin{figure}
\begin{center}
\begin{overpic}[width=.75\textwidth]{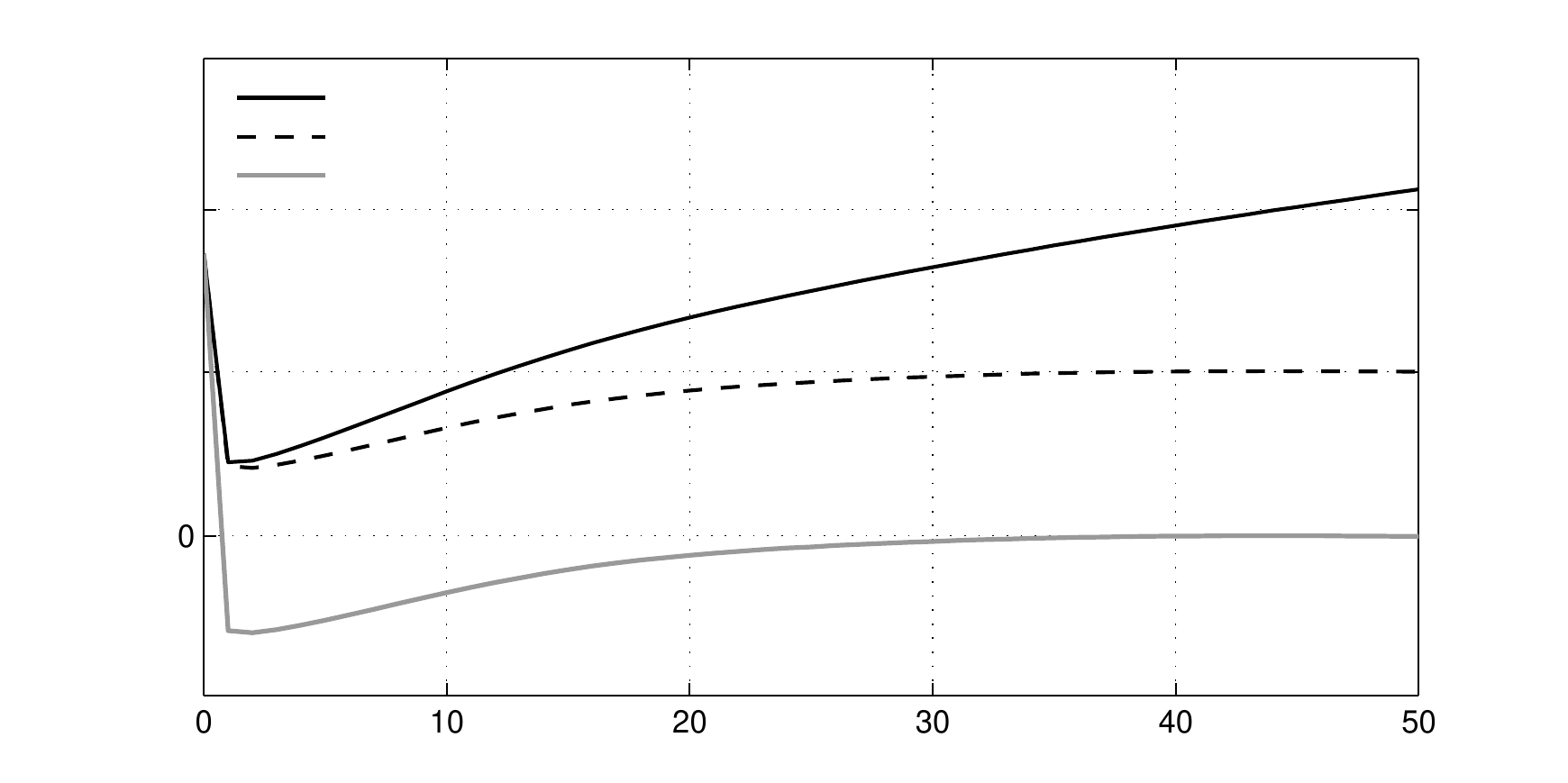}
\put(42,0){Parameter index, $i$}
\put(2,14){\begin{sideways}Markov parameters, \normalfont{$\mathcal{H}_i$}\end{sideways}}
\put(21,43){{\small$\mathcal{H}_i$} from OKID}
\put(21,40.5){{\small$\mathcal{H}_i-C_{\alpha}\alpha$} }
\put(21,38){{\small$\mathcal{H}_i-C_{\alpha}\alpha-C_{\dot\alpha}\dot\alpha$}}
\put(14,36.5){\framebox(27,9){}}
\put(91,15.5){\begin{sideways}$\underbrace{~~~~~~~~~~~~~}_{}$\end{sideways}}
\put(93,20){$C_{\dot\alpha}$}
\put(11,33){\begin{rotate}{270}$\underbrace{~~~~~~~~~~~~~~~~~~~~}_{}$\end{rotate}}
\put(6,24){$C_{\ddot\alpha}$}
\put(59.5,33){\rule{1.2pt}{2.5mm}} 
\put(59.5,34.5){\begin{sideways}\rule{1.2pt}{14mm}\end{sideways}} 
\put(62,35.5){$C_{\alpha}$}
\end{overpic}
\end{center}
\vskip -.1in
\caption{Markov parameters from OKID using Gaussian white noise input for a flat plate at $\alpha_0=15^{\circ}$ and $\Rey=100$.}
\label{fig:markov}
\end{figure}

Figure~\ref{fig:markov} shows the Markov parameters from OKID, which are the discrete-time impulse response parameters in $\ddot\alpha$ for a linearized system of the form in Eq.~(\ref{eq:rom3}).    Because an impulse in $\ddot\alpha$ is a step in $\dot\alpha$ and a ramp in $\alpha$, there is a linear growth in the lift coefficient as the angle of attack increases linearly with time.  Clearly it is not possible to obtain the impulse response of the linearized system from a step in $\dot\alpha$ in a real experiment, since the linear growth in $\alpha$ would quickly excite nonlinear phenomena; this is one motivation for the OKID method, which estimates the linear impulse response from a bounded maneuver in a region where the linear approximation is valid.  

To identify a model of the form in Eq.~(\ref{eq:rom3}), we first identify the lift coefficient slope $C_{\alpha}$.  After subtracting off $C_{\alpha}\alpha$, $C_{\dot\alpha}$ is the steady-state value (since this is a step in $\dot\alpha$).  After these modifications, the first Markov parameter is $C_{\ddot\alpha}$, and the remaining parameters are used by the ERA to obtain a low-dimensional, discrete-time model $(\mA,\mB_{\ddot\alpha},\mC)_r$ for the transient dynamics $(\mA,\mB_{\ddot\alpha},\mC)$ in Eq.~(\ref{eq:rom3}).  

% MIMO TRANSFER FUNCTIONS
\subsection{Identifying multi-input, multi-output models with the ERA}\label{chap:models:algorithms:MIMO}
It is possible to extend the methods above to identify models with multiple inputs and multiple outputs (MIMO), such as a model with pitch, plunge, and surge inputs, as in Eqs.~(\ref{eq:MIMO}),~(\ref{eq:MIMO2}), and~(\ref{eq:MIMO3}).  As with the single-input, single-output (SISO) case, this starts with impulse-response data for each separate input, possibly estimated using OKID.  All of the remaining steps in the procedure are identical, including identifying the quasi-steady and added-mass coefficients for each input--output pair, and formatting the remaining dynamics into a sampled discrete-time impulse.  ERA is used to identify a model for the remaining transient dynamics, and so a Hankel matrix is constructed from the MIMO Markov parameters, which each have size $q \times p$, where $q$ is the number of outputs and $p$ is the number of inputs.  ERA will then identify the remaining dynamics as before without a significant increase in model order.

\subsection{Application of the ERA/OKID to systems with added mass}\label{chap:models:algorithms:naiveokid}
One may attempt to identify the entire model in Eq.~(\ref{eq:rom2}) or Eq.~(\ref{eq:rom3}) using ERA directly on the Markov parameters from OKID.  However, simply applying ERA to the Markov parameters from OKID, which correspond to an impulse response in $\ddot\alpha$, will result in an unstable model because of the linear growth in $C_L$.  The transfer function from $\ddot\alpha$ to $C_L$ contains a double integrator, which is unstable.  These Markov parameters are shown in Figure~\ref{fig:markov}.  In addition, if one applies ERA/OKID to the pair $\{\dot\alpha,C_L\}$ instead of $\{\ddot\alpha,C_L\}$, the resulting model will have input $u=\dot\alpha$ and will not capture the correct added-mass forces.  
The transfer function from $\dot{\alpha}$ to $C_L$ is improper, requiring a derivative to capture $\ddot\alpha$ forces.  It is impossible to represent the derivative of an input in state-space, and so approximating the derivative term ($\dot{u}=\ddot\alpha$) will be inaccurate and require high model order.  Therefore, the correct input to the system is $\ddot\alpha$, and the $C_{\alpha}$ and $C_{\ddot\alpha}$ terms must be identified and removed before applying ERA.   

\begin{figure}
\begin{center}
\begin{tabular}{cc}
\begin{overpic}[width=.45\textwidth]{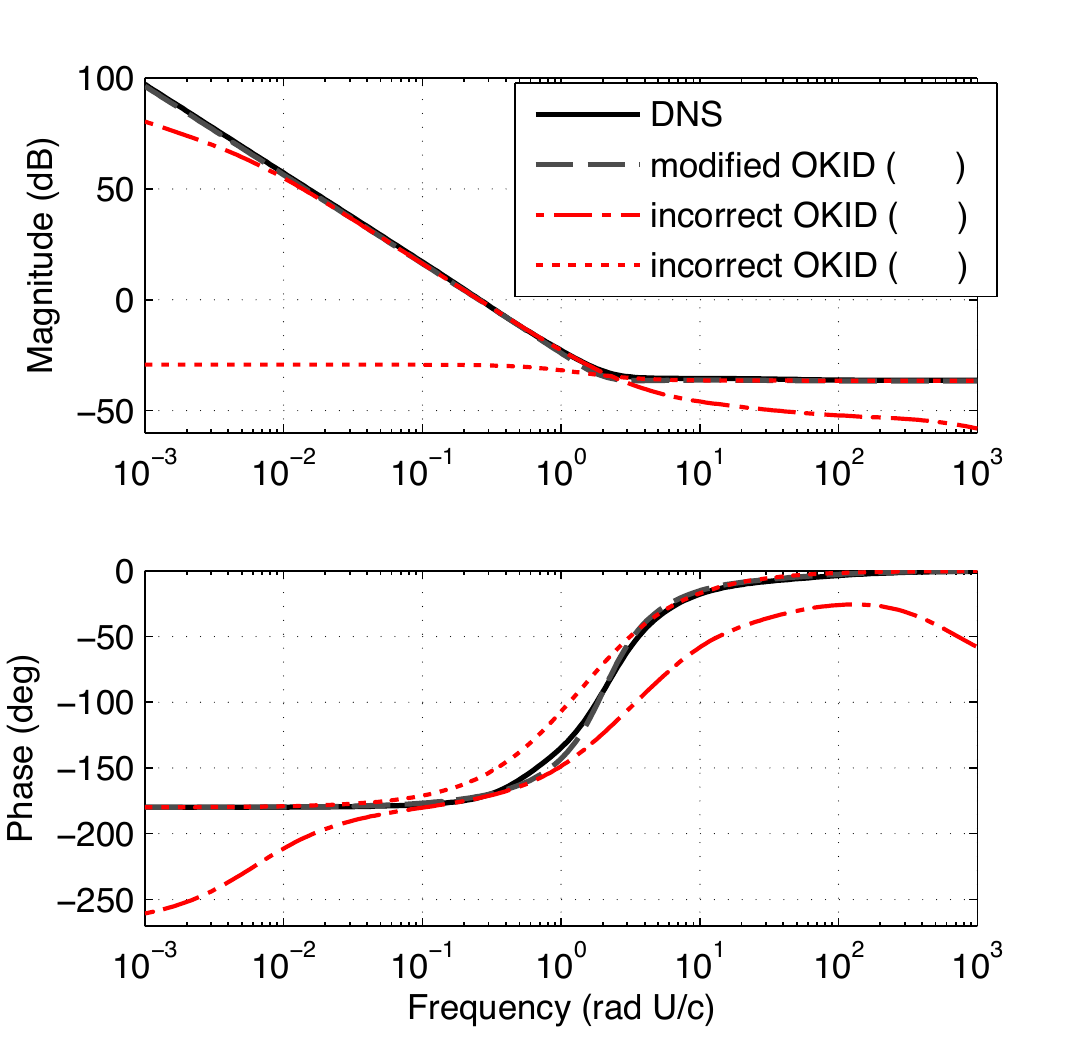}
\put(85,80.5){$\ddot\alpha$}
\put(85,76){$\dot\alpha$}
\put(85,71.5){$\ddot\alpha$}
\put(2,2){(a)}
\end{overpic}
&
\begin{overpic}[width=.45\textwidth]{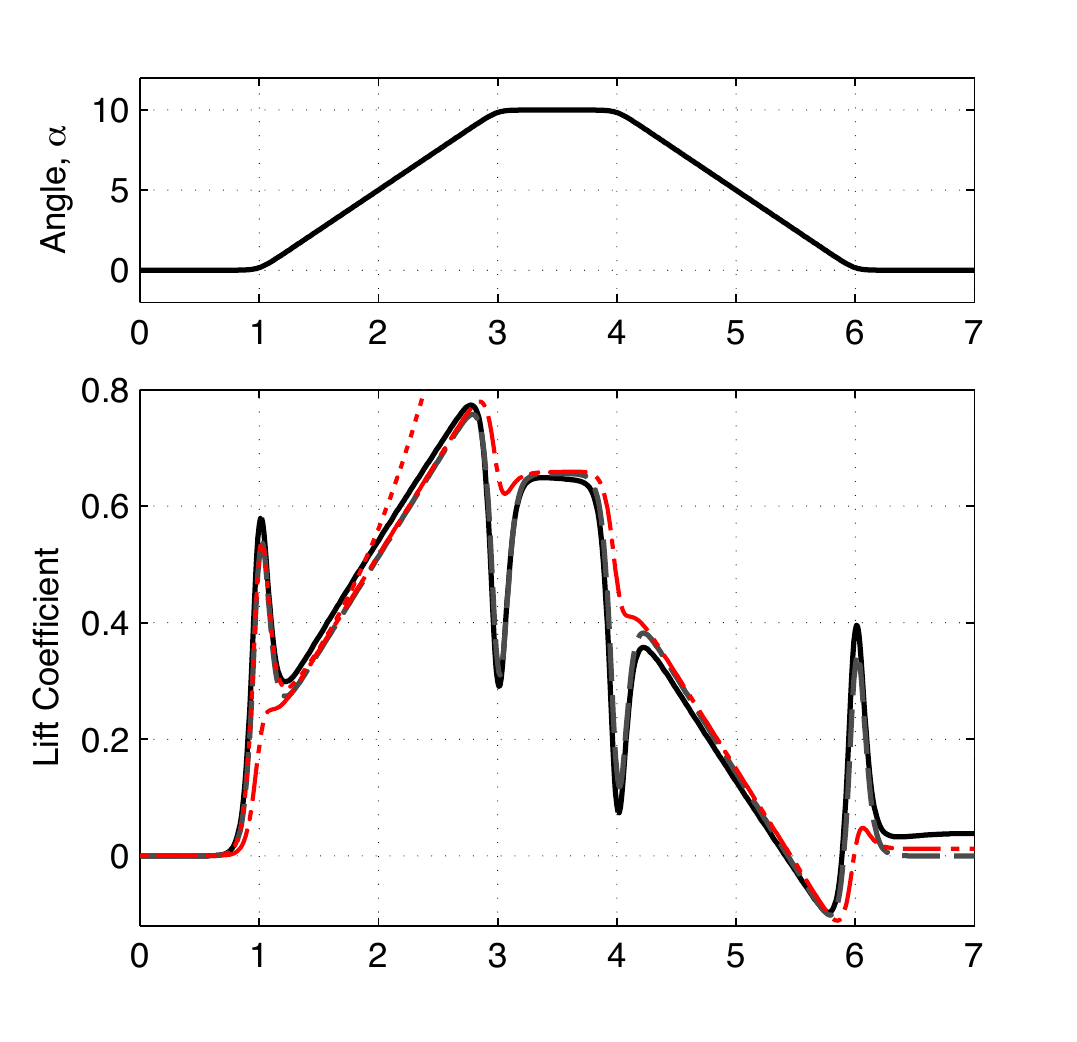}
\put(27,2){{\small{Convective time, $(\tau=tU/c)$}}}
\put(2,2){(b)}
\end{overpic}
\end{tabular}
\end{center}
\vskip -.2in
\caption{Correct and incorrect use of OKID to model pitch dynamics.  (a) Bode plot, and (b) pitch-up, hold, pitch-down maneuver.}
\label{fig:failB}
\end{figure}
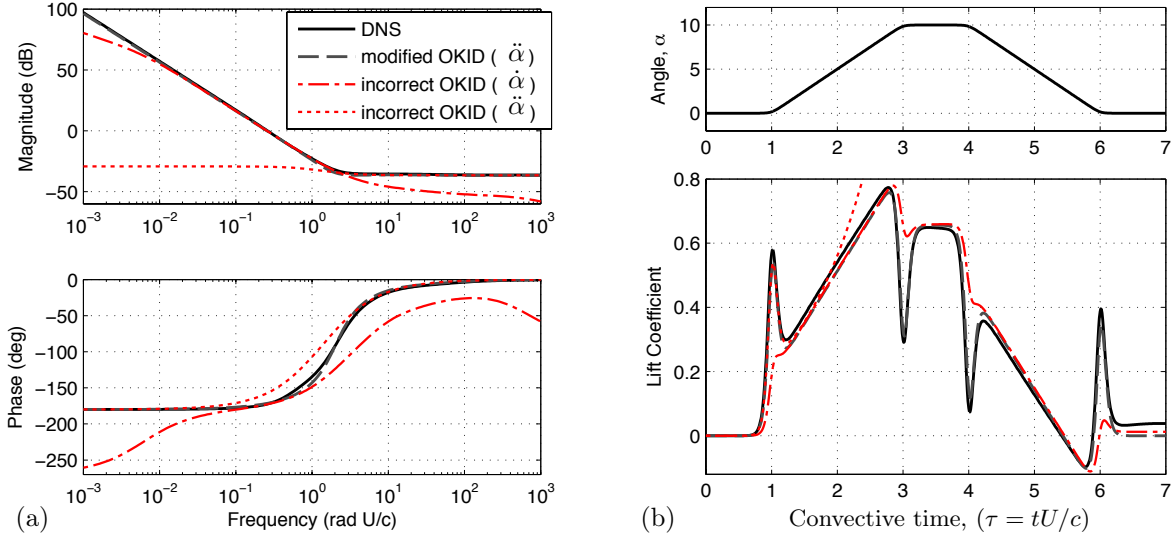

Examples of these systematic failures are shown in Figure~\ref{fig:failB}.  The correct model based on the method in Section~\ref{chap:models:algorithms:3} is labeled ``{modified OKID ($\ddot\alpha$)}''.  The two incorrect uses of ERA/OKID are labeled ``{incorrect OKID}.''  Direct numerical simulation at $\Rey=100$ (see Section~\ref{sec:results}) are labeled ``DNS.''  The incorrect OKID model with input~$\ddot\alpha$ is unstable, as seen in Figure~\ref{fig:failB}~(b).  The incorrect OKID model with input $\dot\alpha$ is unable to capture the added-mass peaks in Figure~\ref{fig:failB}~(b).  Increasing the model order does not significantly improve this behavior.  However, the modified OKID model is stable and is accurate at all frequencies (nearly indistinguishable from DNS).

%%%%%%%%%%%%%%%%%%%%%%%%
%%% RESULTS
%%%%%%%%%%%%%%%%%%%%%%%%
\section{Results}\label{sec:results}
The multi-input models and modeling procedures of the previous sections are demonstrated on a pitching and plunging flat plate at Reynolds number 100 using direct numerical simulations.  In addition, robust controllers are developed, based on reduced-order models for pitch, to track reference lift coefficients at various angles of attack.  The controllers accurately track a reference lift despite significant sensor noise and nonlinear flow separation.

The numerical results in this paper are based on direct numerical simulations (DNS) of the incompressible two-dimensional Navier-Stokes equations using the fast multi-domain immersed boundary projection method (IBPM) of~\cite{taira:07ibfs,taira:fastIBPM}.  The boundary conditions are a specified velocity $\m{v}_B$ at points on the immersed body and uniform flow in the far field, which is valid for a large enough domain.  An efficient multi-domain approach is used, consisting of a series of nested grids, each twice as large as the last.  The Poisson equation is solved on the largest grid with uniform flow boundary conditions.  Next, the interpolated stream function from the larger grids are used as boundary conditions for the Poisson equation on the next smaller grid. This method has been rigorously validated in two-dimensions, as well as in three-dimensions against an oil tow-tank experiment at $\Rey=100$ for a flat rectangular wing with low aspect ratio ($\text{AR}=2$) at $\alpha=30^{\circ}$.

\begin{figure}
\begin{center}
\begin{tabular}{cc}
\begin{overpic}[width=.475\textwidth]{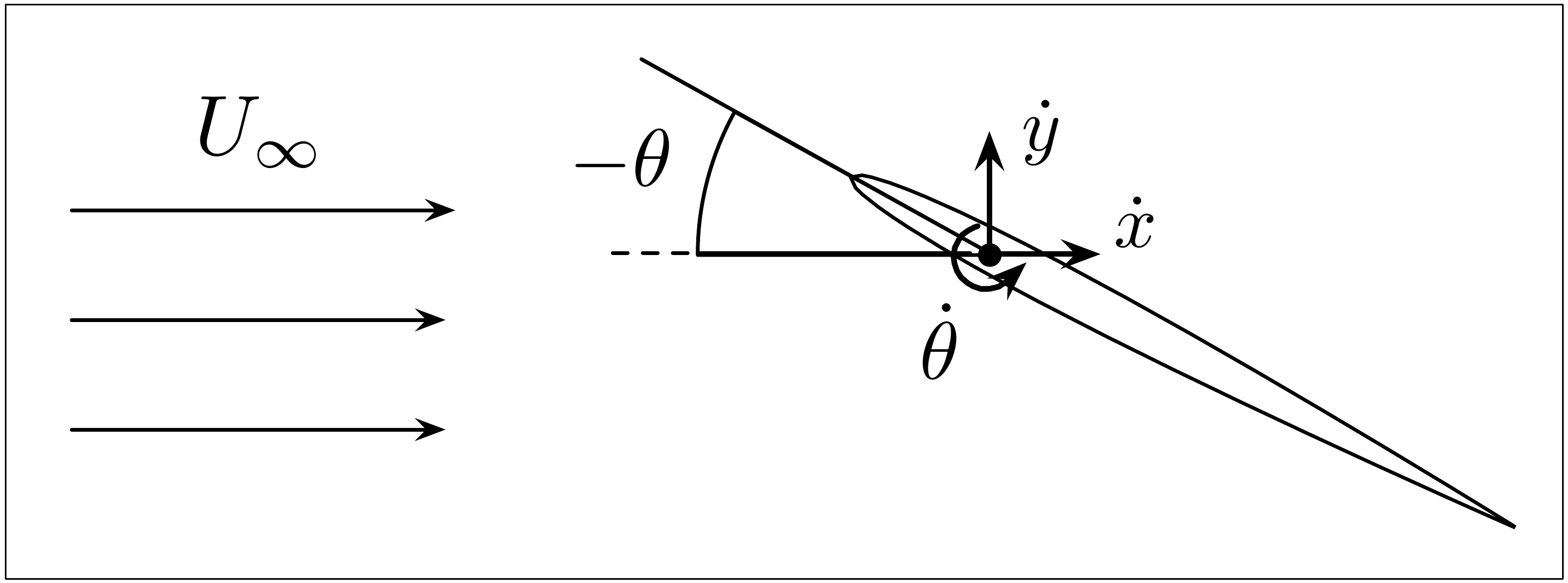}
\put(2,3){(a)}
\end{overpic}&
\begin{overpic}[width=.475\textwidth]{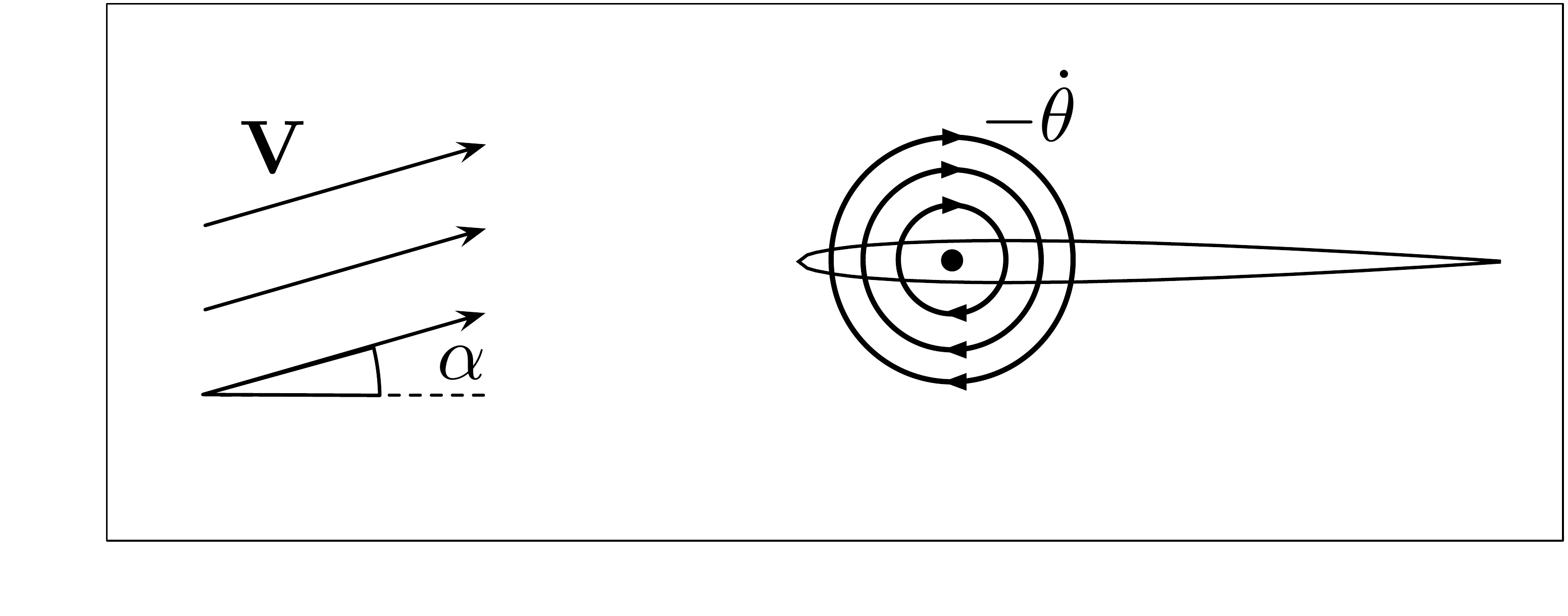}
\put(2,3){(b)}
\end{overpic}
\end{tabular}
\vskip -.05in
\caption{Illustration of wing motion (a) in the wind-tunnel frame and (b) in the wing-fixed frame.}
\vskip -.25in
\label{fig:UBF}
\end{center}
\end{figure}

Instead of solving the equations of motion in the wind-tunnel frame, we solve them in the body-fixed frame of the wing by introducing a moving base flow with uniform and purely rotational components.  The wing kinematics in each frame is shown in Figure~\ref{fig:UBF}.  In the body-fixed frame, boundary points are fixed relative to the grid regardless of wing motion.  Therefore, the matrix involved in solving for the boundary forces does not change in time and can be decomposed by a Cholesky factorization once at the beginning of the simulation.  This is faster and more accurate than iterative methods when the boundary points on the wing move relative to the grid.

The computational domain consists of five nested grids, the finest covering a domain of $4c\times 4c$ and the coarsest covering a domain of $64c\times 64c$, where $c$ is the chord length of the plate.  Each grid has  resolution of $400\times 400$, which is sufficient for converged results. 

\subsection{Reduced-order models for flat plate at $\Rey=100$}
\begin{figure}
\begin{center}
\vskip -.275in
%\hspace{-.5in}
\begin{overpic}[width=.7\textwidth]{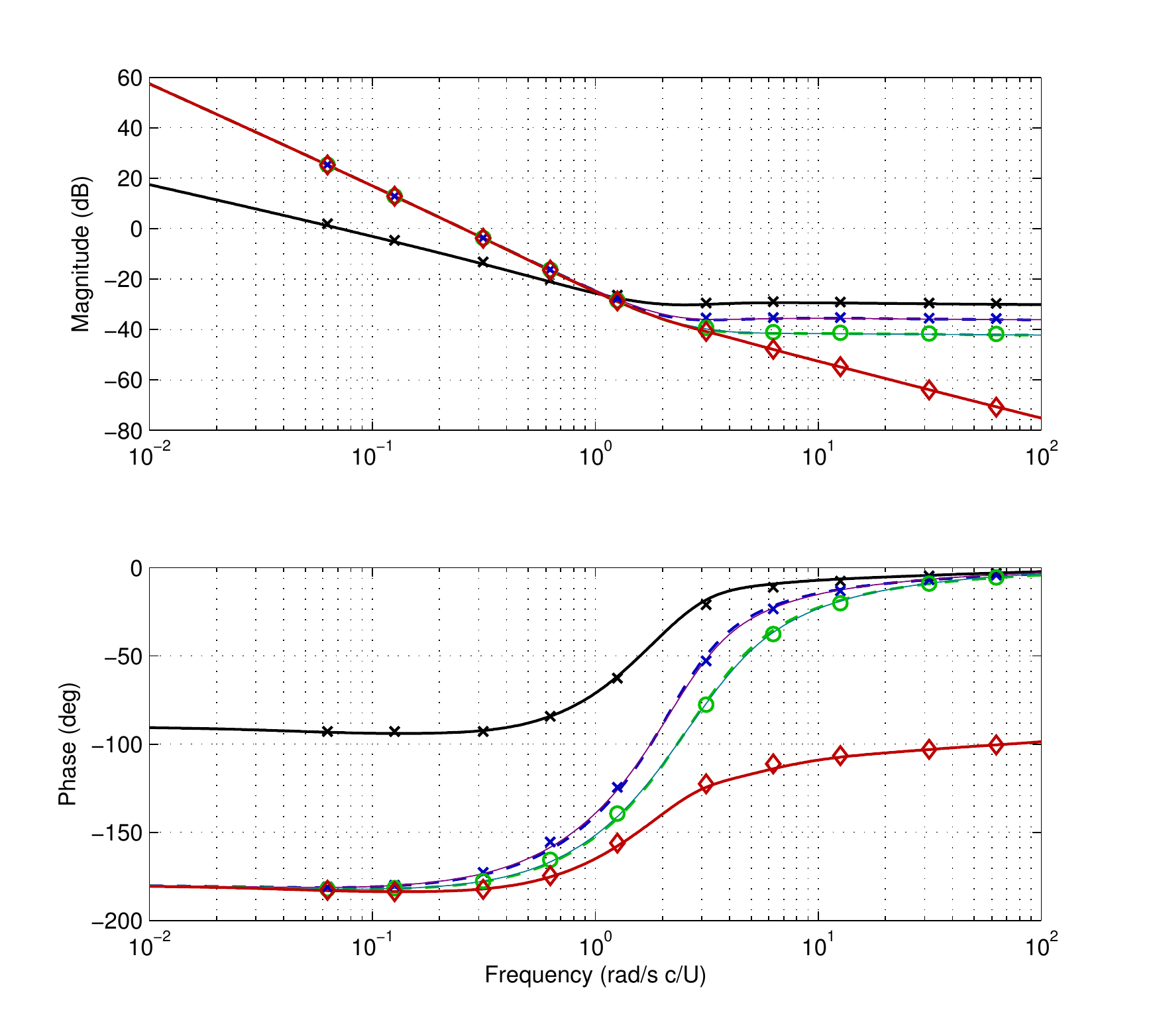}
{\small
\put(91,63){\begin{rotate}{20}$\leftarrow$\end{rotate}}
\put(93,64.5){ plunge}
\put(91,61.5){$\leftarrow\,$leading-edge}
\put(91,60){\begin{rotate}{-20}$\leftarrow$\end{rotate}}
\put(93,58.5){ quarter-chord}
\put(91,53){\begin{rotate}{-20}$\leftarrow$\end{rotate}}
\put(93,51.5){ middle-chord}

\put(21,28){plunge}
\put(36.5,22){leading-edge $\rightarrow$}
\put(62.5,28){$\leftarrow$ quarter-chord}
\put(74,21.5){middle-chord}}
%\put(28.5,59.5){{\small(PL)}}
%\put(27.5,57){{\small$a=-0.5$ (LE)}}
%\put(27.5,54){{\small$a=-0.5$ (MC+PL/2)}}
%\put(27.5,51){{\small$a=-0.25$ (QC)}}
%\put(27.5,48){{\small$a=-0.25$ (MC+PL/4)}}
%\put(27.5,45){{\small$a=0.0$ (MC)}}
%\put(31.2,39.4){\small{, $a=-0.5$}}
%\put(31.2,36.4){\small{, $a=-0.25$}}
%\put(31.2,33.6){\small{, $a=0.0$}}
\end{overpic}
\vskip -.275in
\caption{Bode plot showing that pitch about leading-edge and quarter-chord points is linear combination of pitching about the middle-chord and plunging motion.  Solid lines indicate a single-input model, dashed lines indicate a linear combination of plunge and mid-chord pitching according to Eq.~(\ref{eq:MIMO2}), and symbols indicate data from single-frequency DNS.  ERA model order is $r=7$ for all cases.}\label{fig:pitchpoint}
%\end{center}
%\end{figure}
%
%\begin{figure}
%\begin{center}
\begin{overpic}[width=.7\textwidth]{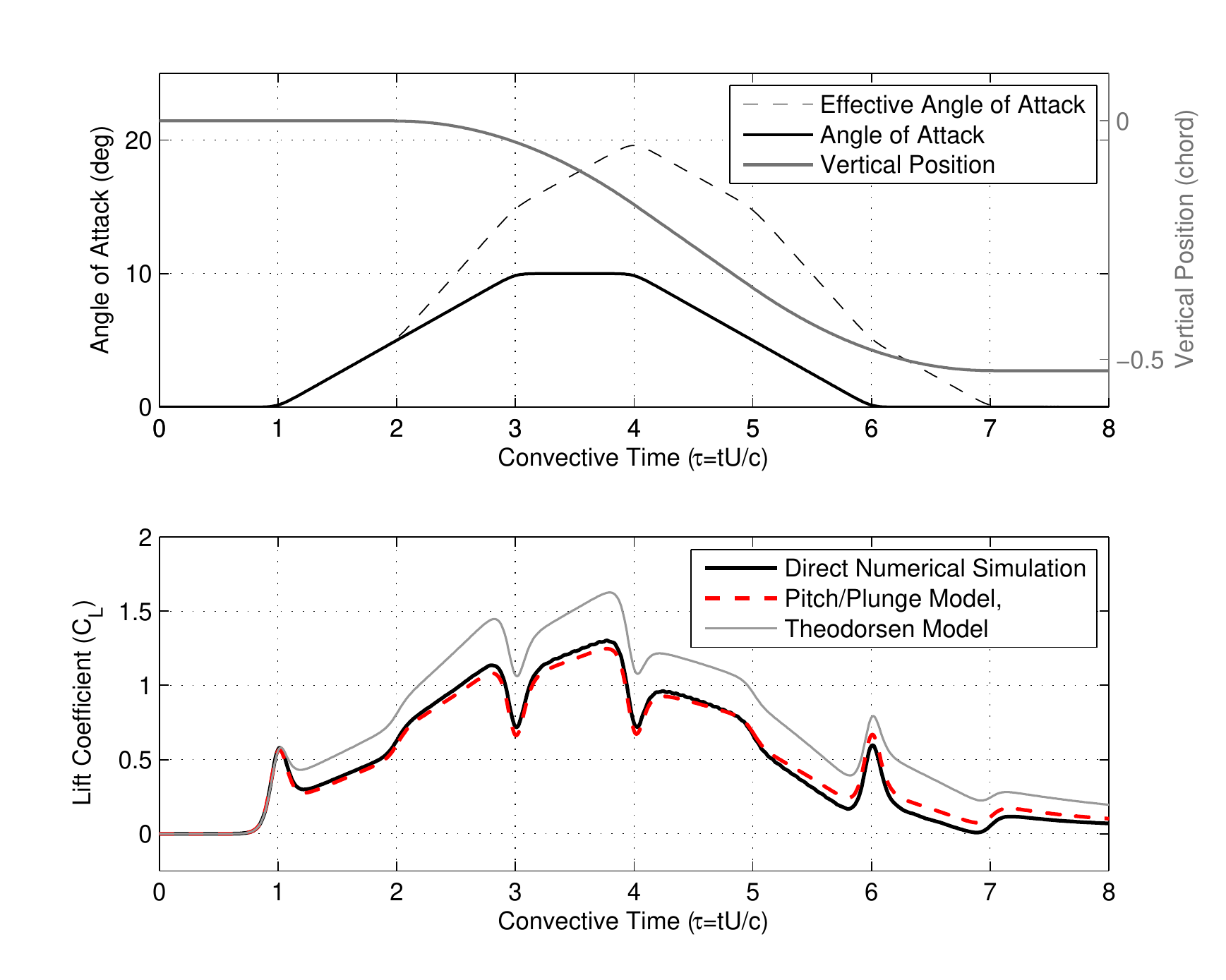}
\put(5,5){(b)}
\put(5,42){(a)}
\end{overpic}
\vskip -.22in
\caption{Combined pitch/plunge maneuver.  (a) The angle-of-attack
  and vertical center of mass motions.  (b) Lift coefficient
  response is shown for DNS, reduced-order model and Theodorsen's model.}
\label{fig:combined}
\end{center}
\end{figure}

We develop models for plunge and pitch about different pitch-axis locations at various base angles of attack, $\alpha_0$.  At $\alpha_0=0^{\circ}$, we construct a multiple-input model, as in Eq.~(\ref{eq:MIMO2}), for pitch and plunge motion that is parametrized by pitch-axis location $a$.  In particular, pitching motion about any point may be considered equivalent to pitching about the mid-chord in addition to an induced plunge motion.  This is \emph{kinematically} exact, although the fluid dynamic response is not necessarily the sum of the individual mid-chord pitching response and plunging response; this only holds if the responses are \emph{linear} so that superposition applies.  Having identified models for pitching at the mid-chord and plunging, it is possible to reconstruct the model for pitch about the leading-edge and quarter-chord based on a linear combination of these two models.  The model agreement is nearly exact, as shown in Figure~\ref{fig:pitchpoint}, because the responses are approximately linear and superposition holds.  This is similar to the pitch-point parameterization in Theodorsen's model, except that we identify separate models for pitch and plunge dynamics.

The models in Eq.~(\ref{eq:rom2}) and~(\ref{eq:MIMO2}) agree well with the indicial response model and DNS for small-amplitude maneuvers, and so we try maneuvers with a larger amplitude.  Because both the reduced-order model and the indicial response method are linear, they should continue to agree with each other for large amplitude motions, even if they disagree with DNS.  However, the actual flow physics is nonlinear, and so comparison with DNS on a larger maneuver provides a more challenging test case for the models.  Additionally, comparison with Theodorsen's model will highlight the advantages of the reduced-order model.

A large amplitude combined pitch/plunge maneuver is shown in Figure~\ref{fig:combined} (a).  The pitching portion of the maneuver consists of a pitch-up, hold, pitch-down about the leading edge with a maximum angle of $10^{\circ}$.  The plunging portion of the maneuver consists of a step-down in vertical position which is the negative integral of the step-up,
hold, step-down maneuver, and is chosen to have a maximum effective angle of attack based on vertical velocity of $10^{\circ}$.  The motion of $\alpha$ and $\dot{h}$ are given by the following expression for $u$:
\begin{align}
G(t) &= \log\left[\frac{\cosh(b(t-t_1))\cosh(b(t-t_4))}{\cosh(b(t-t_2))\cosh(b(t-t_3))}\right], &    u(t)&= u_{\text{max}}\frac{G(t)}{\max(G(t))};\label{eq:PRHpitch}
\end{align}
where $b=11$, $\alpha_{\text{max}}=10^{\circ}$ and $\dot{h}_{\text{max}}=-0.1745$, which corresponds to $\alpha_e=10^{\circ}$.  For the pitching motion, $t_1=1, t_2=3, t_3=4, t_4=6$, and for the plunging motion, $t_1=2, t_2=4, t_3=5, t_4=7$.  These maneuvers are based on the canonical pitch-up, hold, pitch-down maneuver of~\cite{eldredge:2009,canonical:2010}.

Figure~\ref{fig:combined} shows the performance of each model on the combined pitch/plunge maneuver (a). The bottom plot shows the lift coefficient of each model throughout the maneuver.  The combined pitch/plunge model agrees with DNS, and outperforms Theodorsen's model, which over-predicts the lift throughout due to the idealized $2\pi$ lift slope.  The indicial response model agrees with the reduced-order model (\ref{eq:rom2}), although it is not plotted.  

The frequency response for the reduced-order model at a base angle of $\alpha_0=20^{\circ}$ is shown in Figure~\ref{fig:bode}.  The lift coefficient slope decreases as the angle of attack increases, as observed by the decrease in low-frequency asymptote.  In addition, the model takes longer to equilibrate to motion or disturbances at low frequencies since the low-frequency phase converges to its asymptote at a lower frequency.  There is a strong resonance between $1$ and $10$ rad/s c/u corresponding to a mode that will become unstable and lead to vortex shedding~\cite{brunton:2012b}.

\subsection{Feedback control of unsteady lift coefficient by varying pitch angle}
\begin{figure}
\begin{center}
\begin{overpic}[width=.9\textwidth]{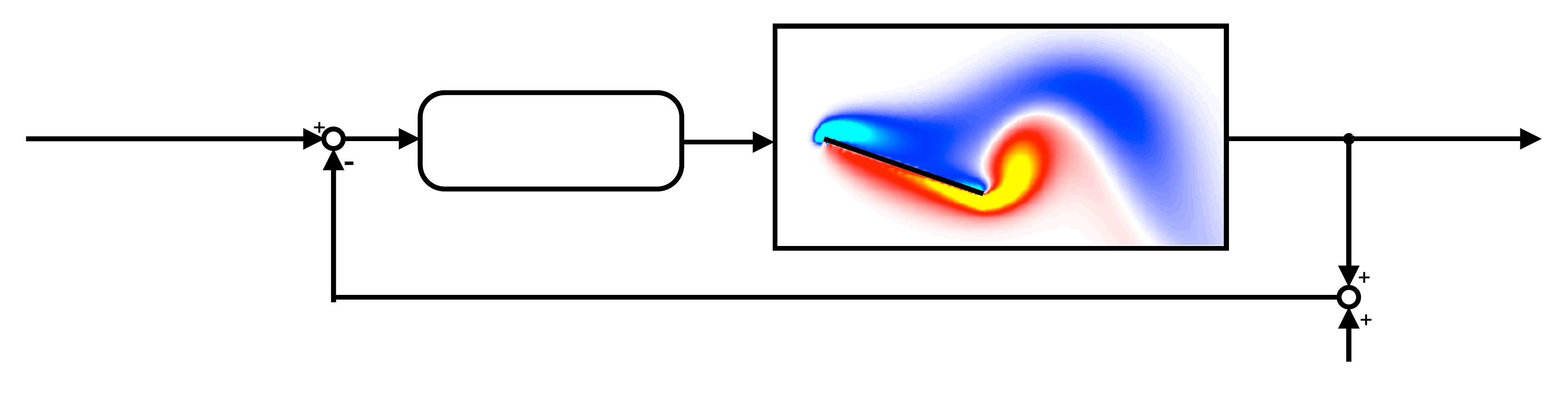}
\put(1,19){Reference Lift, $r_L$}
\put(23,19){$e$}
\put(45,19){$\ddot\alpha$}
\put(79,19){\small{Measured Lift}, $C_L$}
\put(30.,17.25){\small{Controller}}
\put(50.5,23.25){\small{Unsteady Simulations}}
\put(85.5,2.25){$n$}
\end{overpic}
\vskip -.265in
\caption{Control diagram for reference lift tracking in simulations.}\label{fig:controllerscheme}
\vspace{ -.275in}
\end{center}
\end{figure}
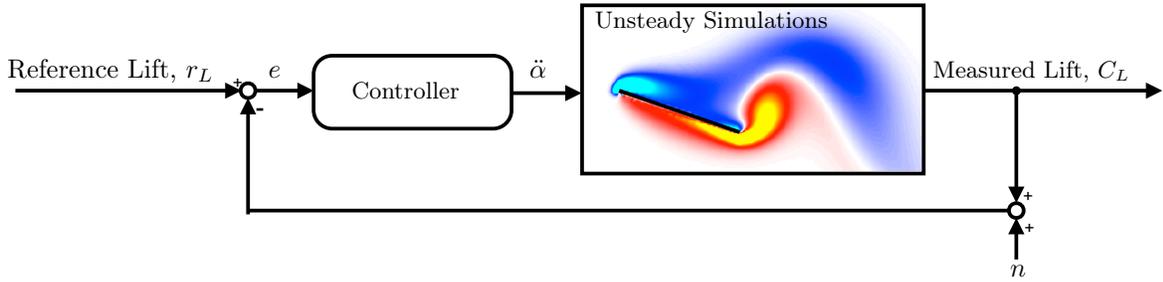

Robust feedback controllers are developed, based on reduced-order models for pitch, to track a reference lift coefficient, $r_L$, while rejecting low-frequency disturbances and attenuating high-frequency sensor noise.  We desire \emph{robust} controllers for unsteady aerodynamic applications for a number of reasons.  First, the actual unsteady fluid dynamics are governed by the nonlinear Navier-Stokes equations, so there are unmodeled dynamics corresponding to the nonlinear terms.  In addition, if control of micro aerial vehicles is the end-goal, it is important to be able to track a reference trajectory despite large gust disturbances and possible environmental disturbances, such as rain or unpredicted changes to the vehicle (damage, change in payload, etc.).  Finally, sensor measurements are inherently noisy, and this is especially a concern at low flow velocities.  The controller schematic is shown in Fig.~\ref{fig:controllerscheme}.

\begin{figure}
\begin{center}
\begin{tabular}{cc}
\begin{overpic}[width=.45\textwidth]{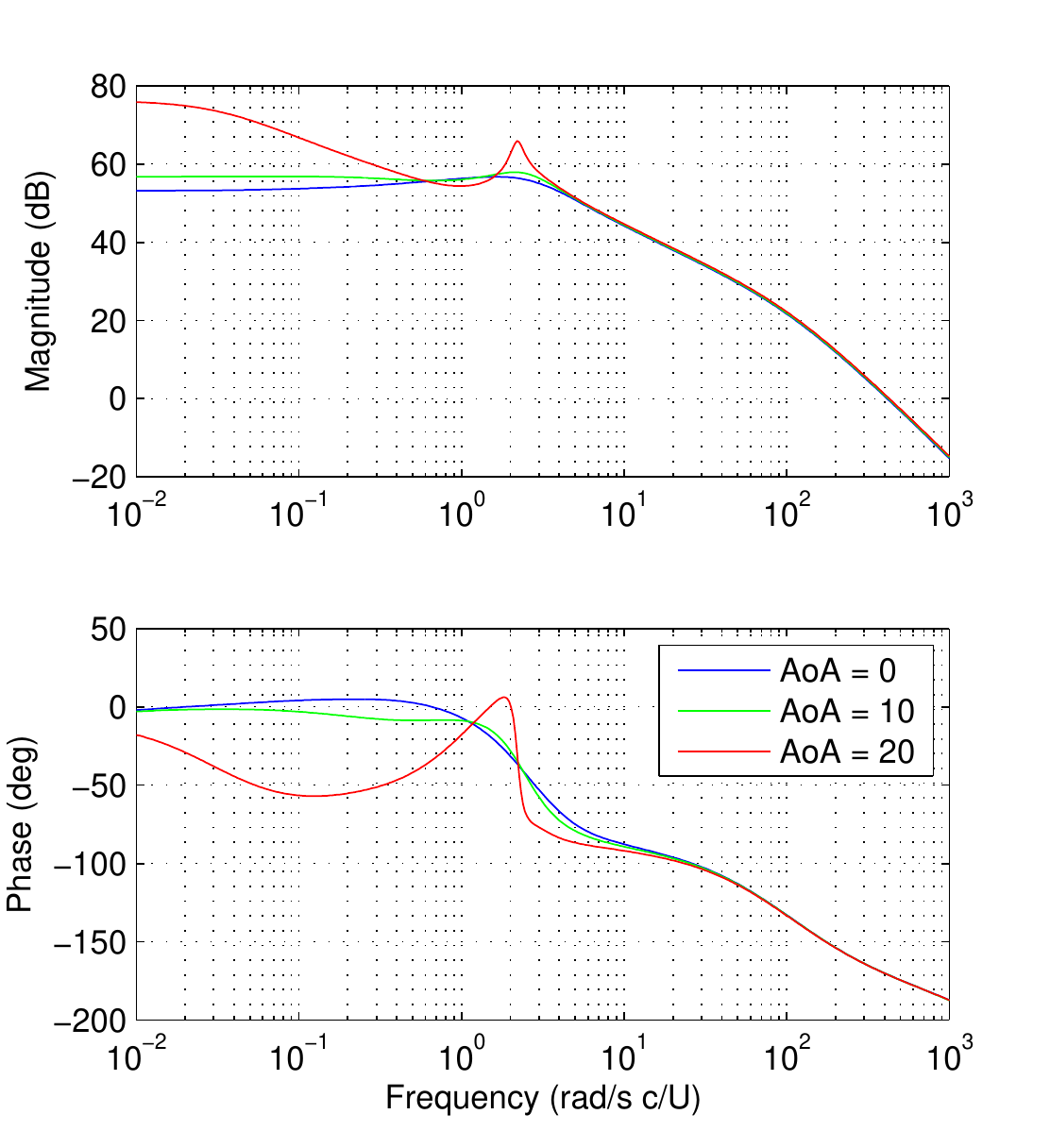} 
\put(0,3){(a)}
\end{overpic}& 
\begin{overpic}[width=.45\textwidth]{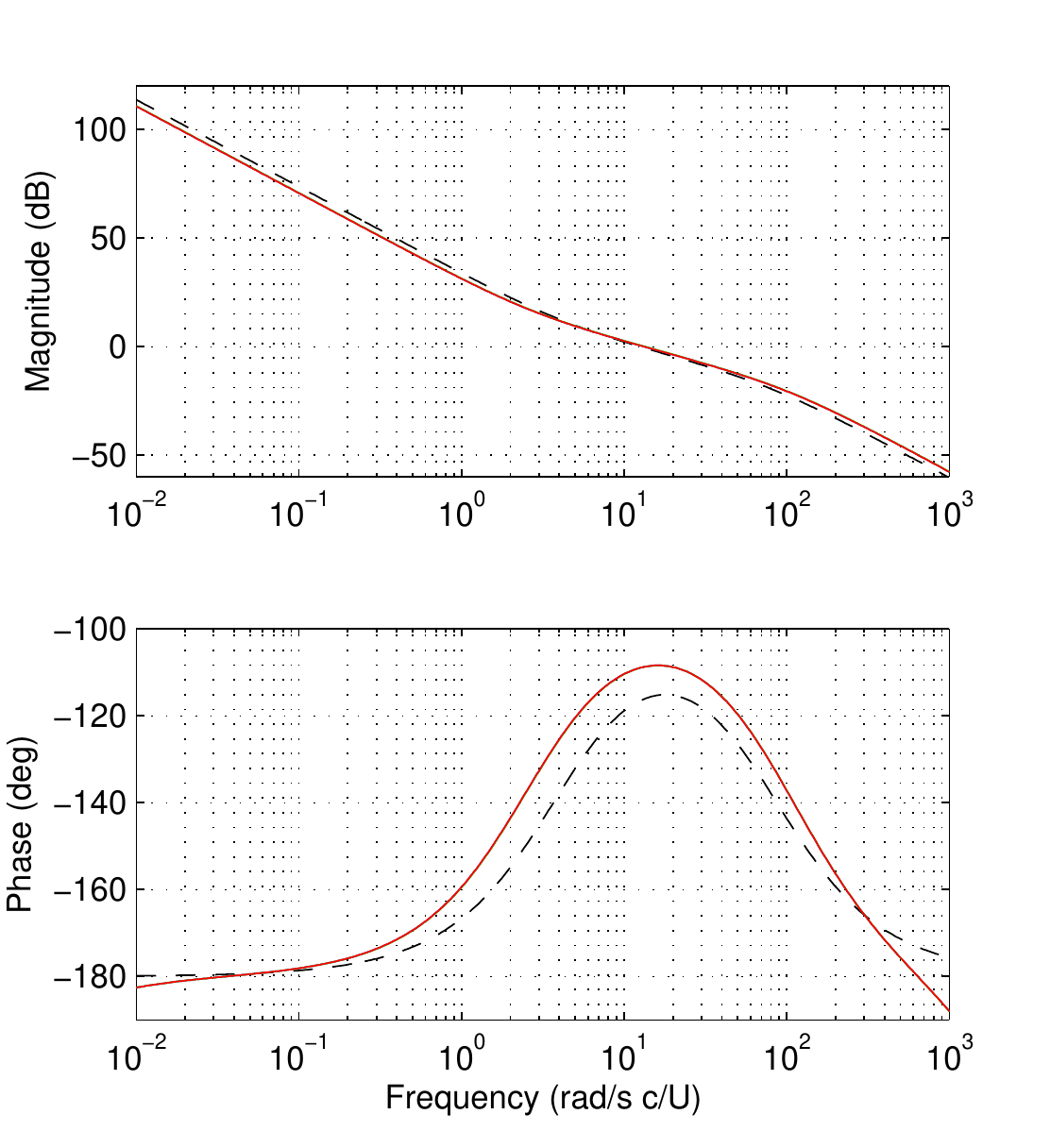} 
\put(0,3){(b)}
\end{overpic}
\end{tabular}
\vspace{-.22in}
\caption{Bode plots for controllers (left) and loop shape (right) for  $\alpha_0=0^{\circ}$, $\alpha_0=10^{\circ}$, and $\alpha_0=20^{\circ}$.  Desired loop shape (right) is given by dashed line.}\label{fig:controllerbode}
\vspace{-.2in}
\end{center}
\end{figure}

Figure~\ref{fig:controllerbode} shows the frequency response of controllers (left) designed for quarter-chord pitching at various base angles of attack using $\mathcal{H}_{\infty}$ loop-shaping~\cite{sp:book}.  The desired loop shape, shown in the plot on the right, is given by $L_d={960(s+4)}/{s^2(s+80)}$.  Actuator roll-off is modeled by $G_a=500/(s+500)$.  

\begin{figure}
\begin{center}
\begin{tabular}{cc}
\begin{overpic}[width=.45\textwidth]{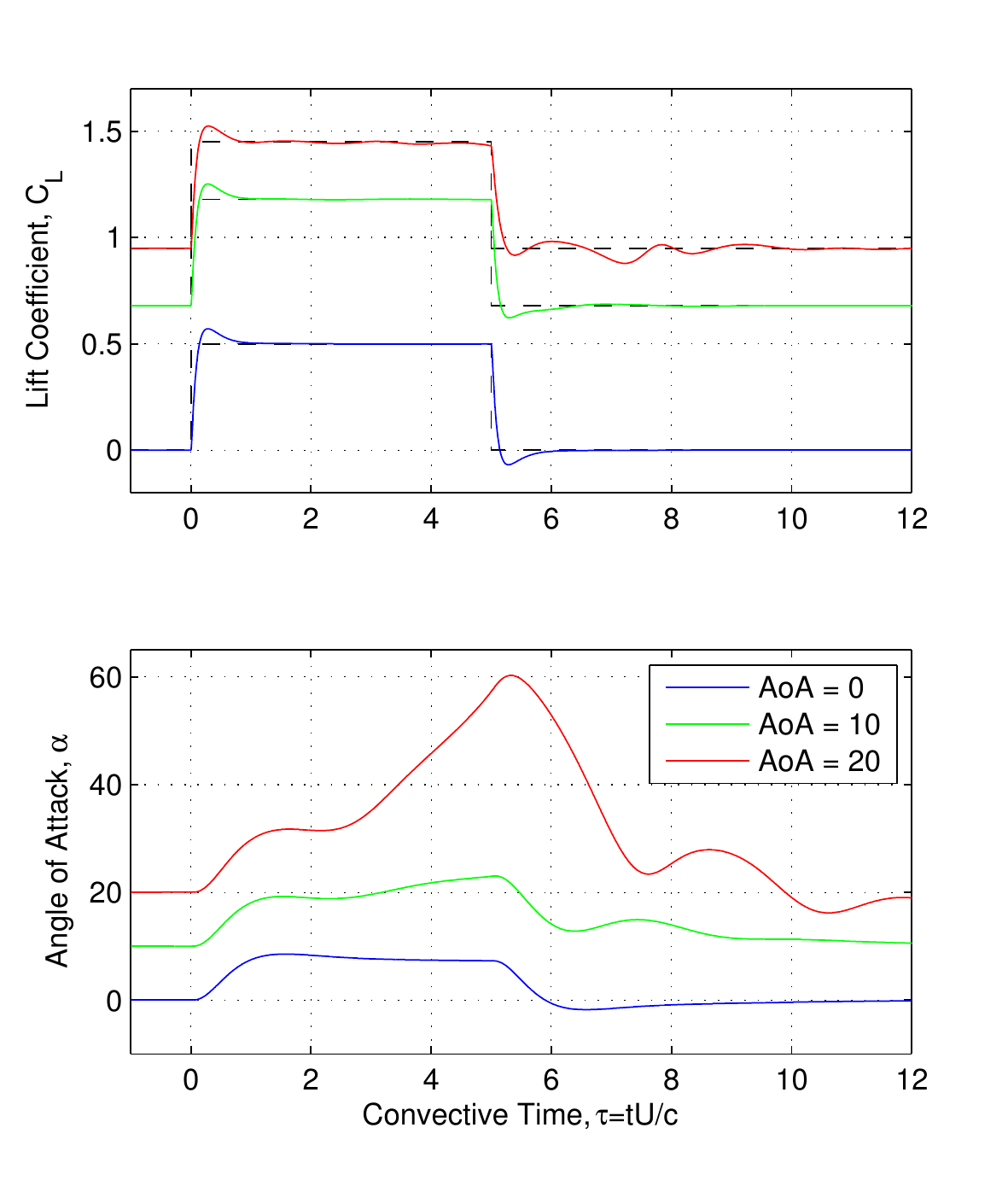} 
% tau = 4.0
\put(35.2,16.8){\Huge$\cdot$}
\put(35.2,23.3){\Huge$\cdot$}
\put(35.2,34.2){\Huge$\cdot$}
\put(36.7,19.6){\footnotesize$A$}
\put(36.7,26.3){\footnotesize$B$}
\put(36.7,34.5){\footnotesize$C$}

\put(35.2,68.7){\Huge$\cdot$}
\put(35.2,81){\Huge$\cdot$}
\put(35.2,85.6){\Huge$\cdot$}
\put(36.7,71.5){\footnotesize$A$}
\put(36.7,80.5){\footnotesize$B$}
\put(36.7,88.8){\footnotesize$C$}

% tau = 8.0
\put(55.4,13){\Huge$\cdot$}
\put(55.4,20){\Huge$\cdot$}
\put(55.4,25.){\Huge$\cdot$}
\put(57,16){\footnotesize $D$}
\put(57,22.5){\footnotesize$E$}
\put(57,28.5){\footnotesize$F$}

\put(55.4,59.7){\Huge$\cdot$}
\put(55.4,72){\Huge$\cdot$}
\put(55.4,77){\Huge$\cdot$}
\put(57,62.5){\footnotesize$D$}
\put(57,71.5){\footnotesize$E$}
\put(57,80){\footnotesize$F$}
\put(-1,6){(a)}
\end{overpic}
& 
\begin{overpic}[width=.45\textwidth]{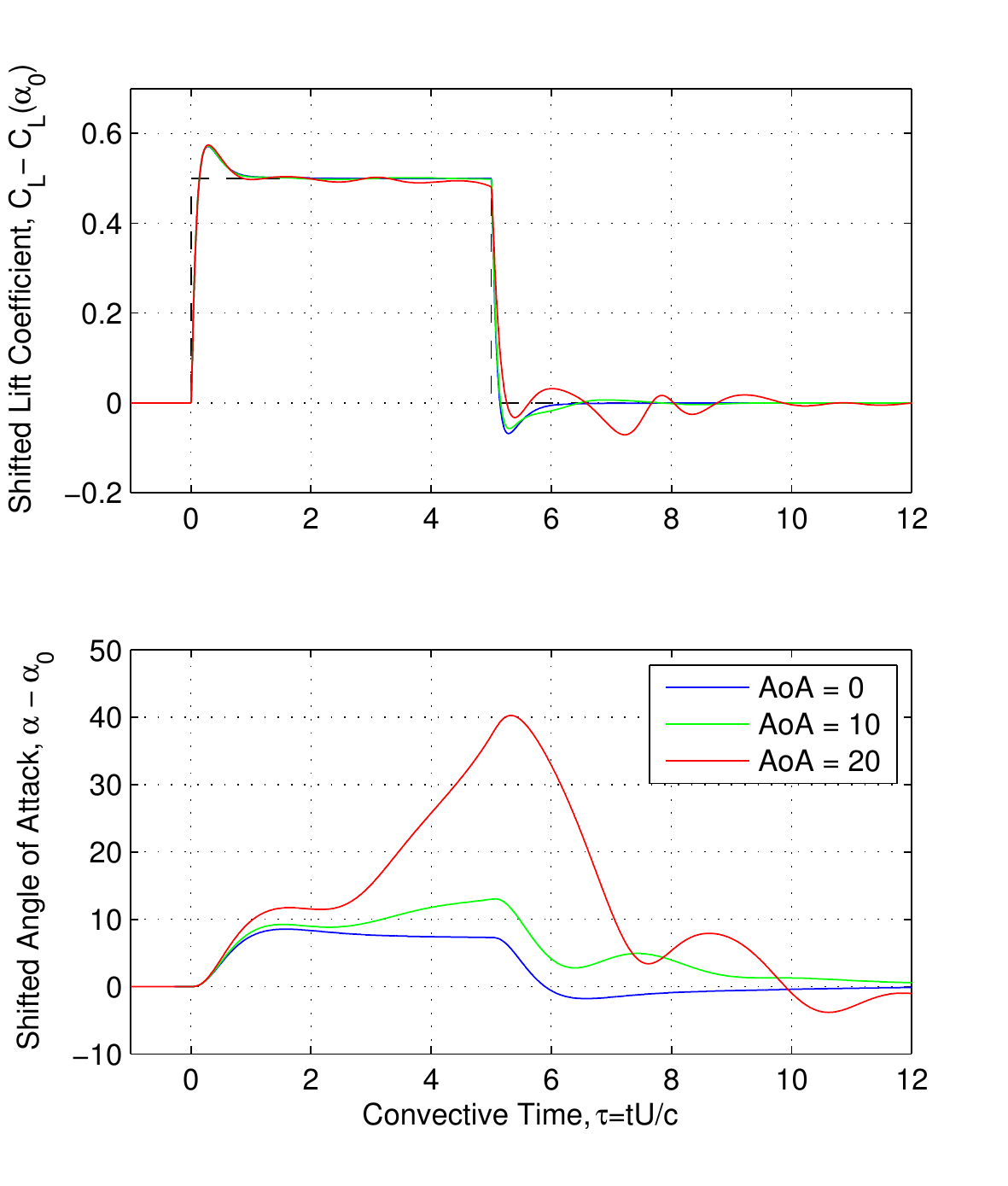} 
\put(-1,6){(b)}
\end{overpic}
\end{tabular}
\begin{tabular}{ccc}
\begin{overpic}[width=.3\textwidth]{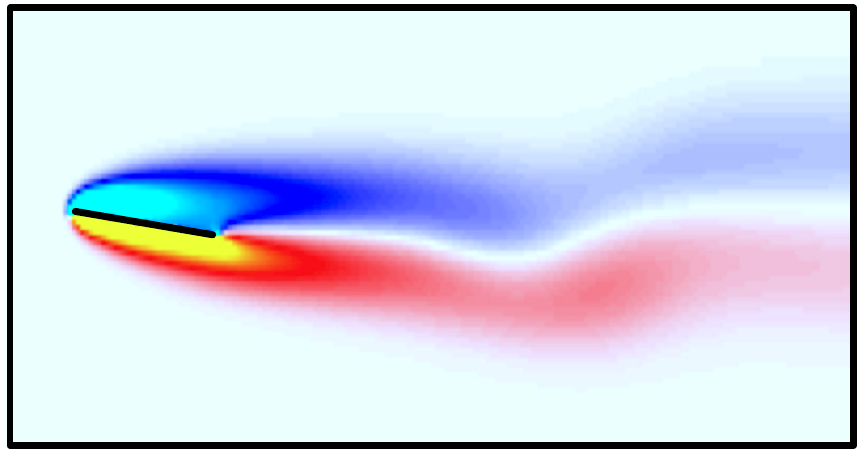} 
\put(5,5){\footnotesize$A$}
\end{overpic}
\begin{overpic}[width=.3\textwidth]{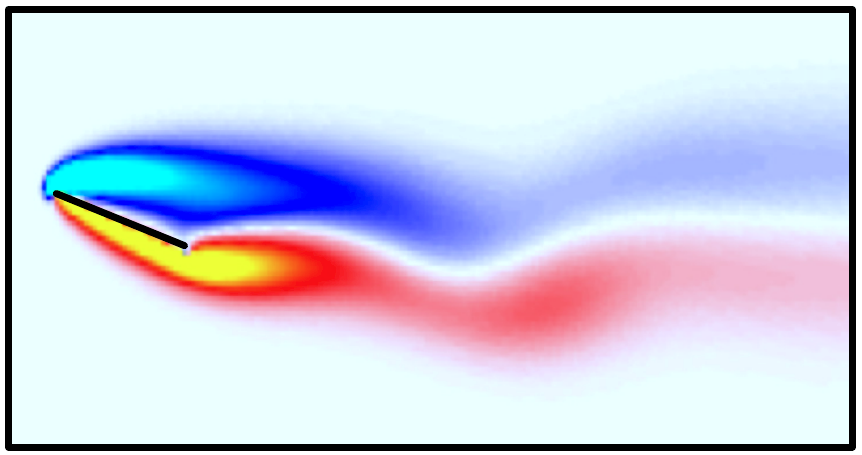} 
\put(5,5){\footnotesize$B$}
\end{overpic}
\begin{overpic}[width=.3\textwidth]{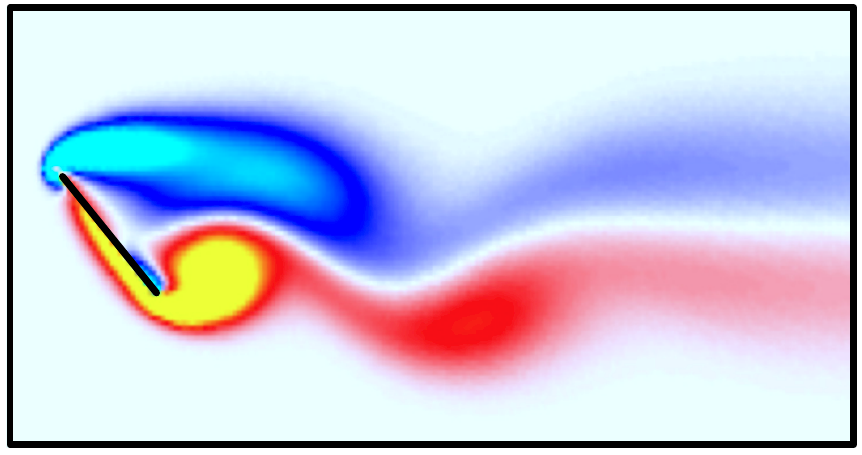} 
\put(5,5){\footnotesize$C$}
\end{overpic}\\
\begin{overpic}[width=.3\textwidth]{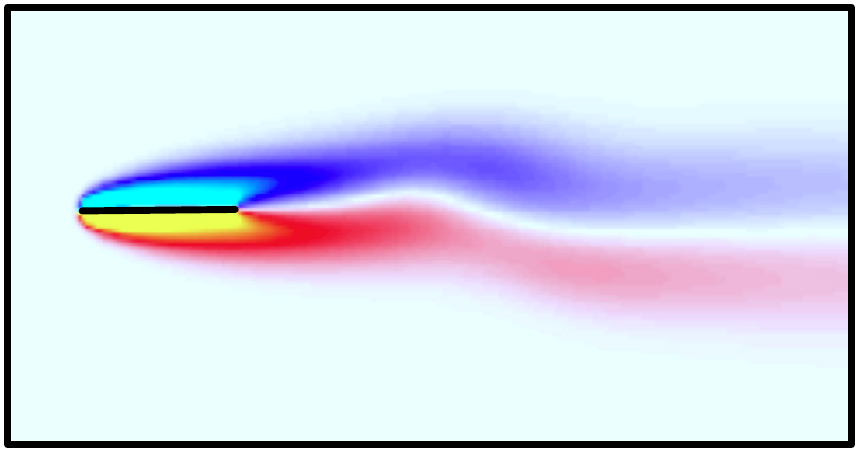} 
\put(5,5){\footnotesize$D$}
\put(-9,4){(c)}
\end{overpic}
\begin{overpic}[width=.3\textwidth]{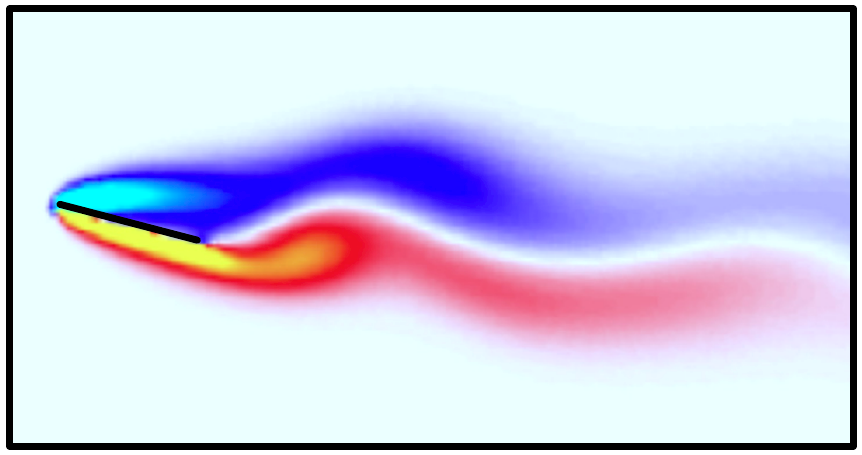} 
\put(5,5){\footnotesize$E$}
\end{overpic}
\begin{overpic}[width=.3\textwidth]{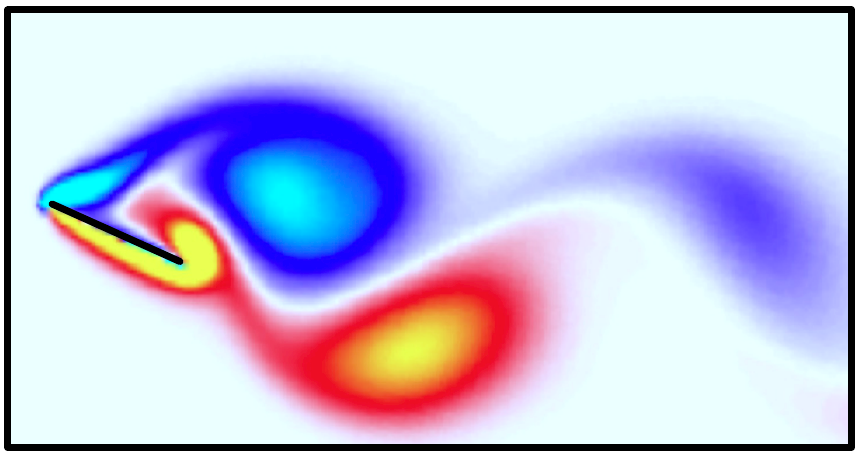} 
\put(5,5){\footnotesize$F$}
\end{overpic}
\end{tabular}
\vspace{-.05in}
\caption{Step response plots from $C_L(\alpha_0)$ to $C_L(\alpha_0)+0.5$ and back down for base angle of attack $\alpha_0=0^{\circ}$, $\alpha_0=10^{\circ}$, and $\alpha_0=20^{\circ}$.  
(a) Unmodified lift coefficient and angle of attack, (b) Shifted lift $C_L-C_L(\alpha_0)$ and angle $\alpha-\alpha_0$ curves.  Reference steps are dashed lines.  (c) Vorticity fields at specific times during maneuver.}\label{fig:controllerstep}
\vspace{-.2in}
\end{center}
\end{figure}
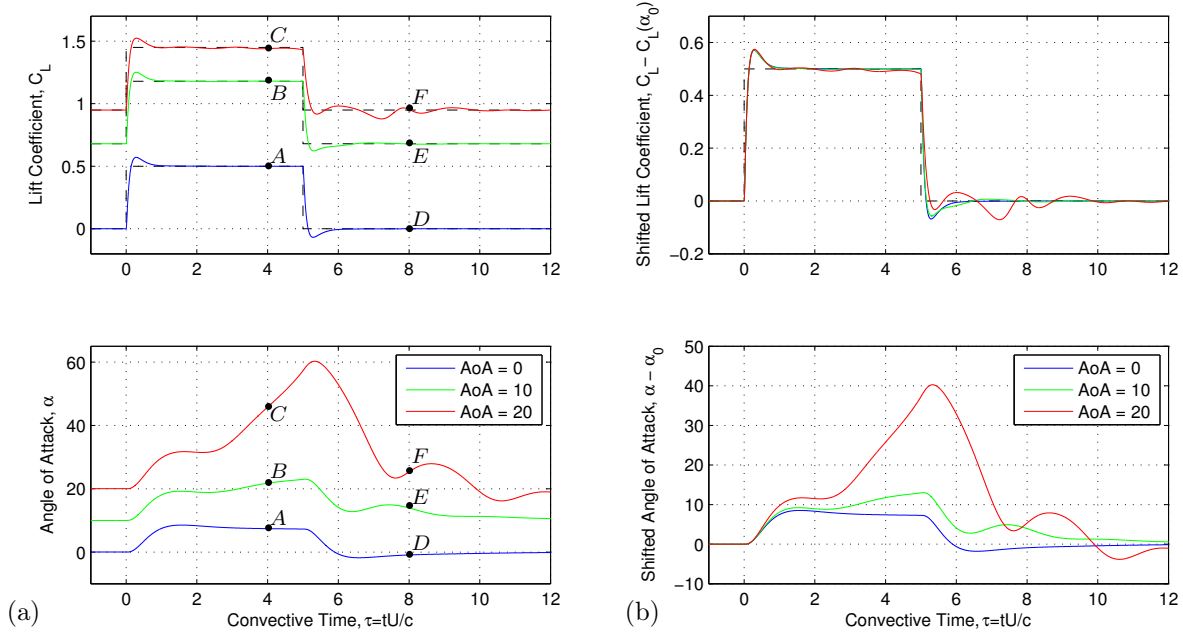

The left panels of Figure~\ref{fig:controllerstep} show the response of the closed-loop system to a commanded step in $C_L$ from $C_L(\alpha_0)$ to $C_L(\alpha_0)+0.5$, followed by a step back down to $C_L(\alpha_0)$ five convective time units later.  The right panels show the lift coefficient and angle of attack, shifted by $C_L(\alpha_0)$ and $\alpha_0$, respectively.  Differences in the lift response and commanded angle of attack indicate the nonlinearity of the system; if the system were truly linear and the models were equivalent, then the angle of attack curves would collapse.  Because of the significantly decreased quasi-steady lift slope at $\alpha_0=20^{\circ}$, it is necessary for the controller to command significantly larger angles of attack to achieve the desired step in lift.  All of the shifted lift coefficient curves collapse during the pitch-up portion of the maneuver, although the lift for the maneuver based at $\alpha_0=20^{\circ}$ deviates after the pitch-down.  For this large $\alpha_0$ case, the transient large angle of attack ($\sim 60^{\circ}$) results in vortex shedding during the step-down.  The subsequent flow is strongly nonlinear and is not well approximated by the $\alpha_0=20^{\circ}$ model.

\begin{figure}
\begin{center}
\begin{overpic}[width=.75\textwidth]{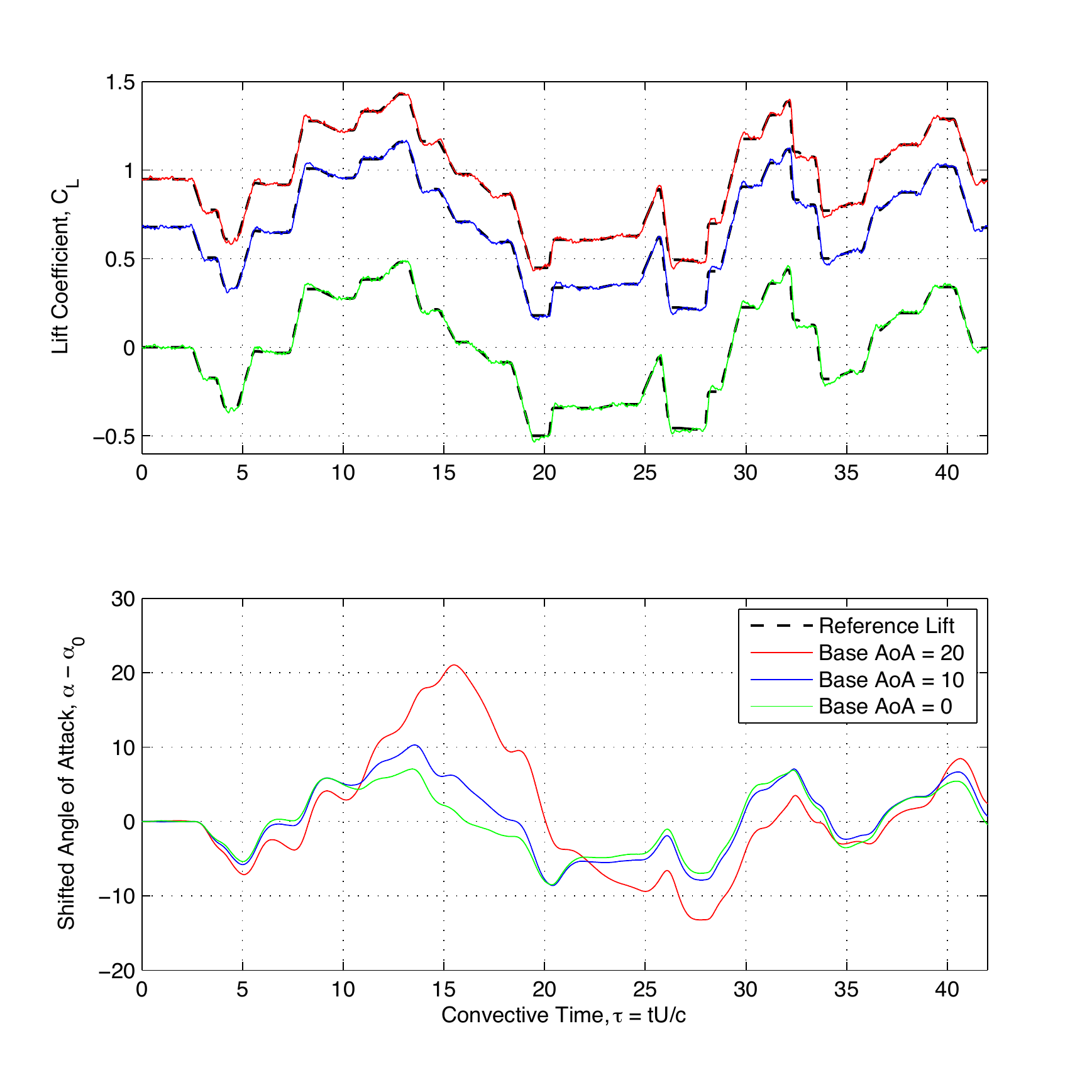}
\end{overpic}
\begin{overpic}[width=.75\textwidth]{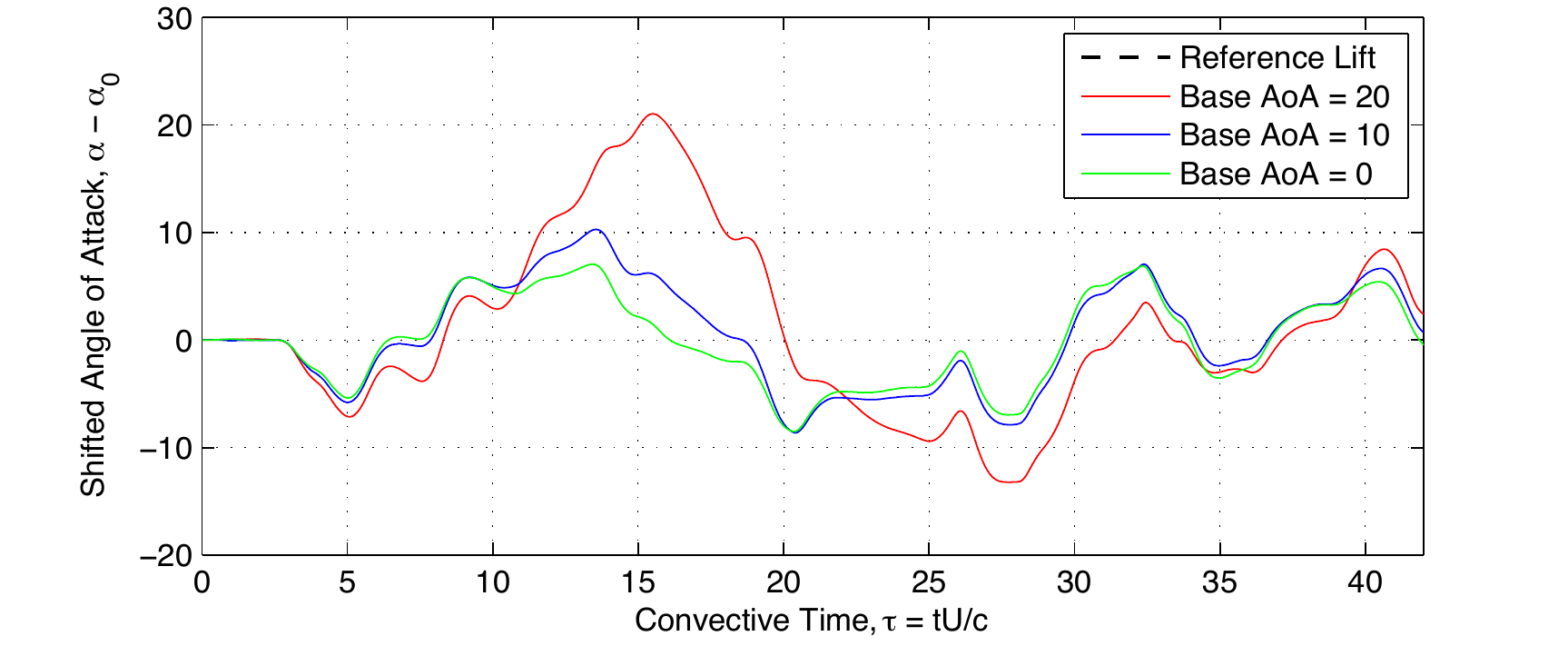}
\end{overpic}
\begin{tabular}{cc}
\begin{overpic}[width=.45\textwidth]{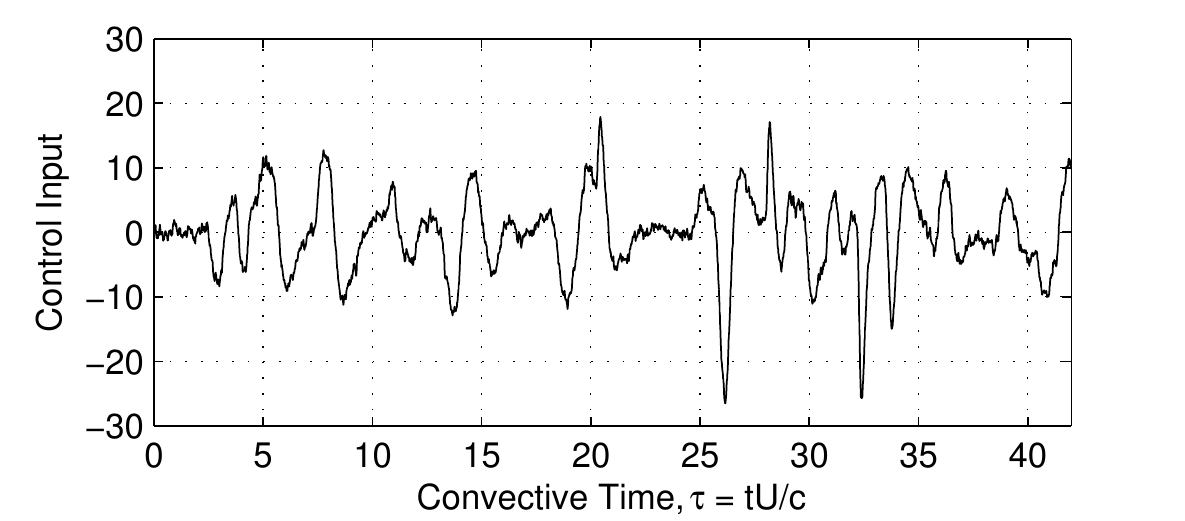} 
\put(2.5,34){\begin{sideways}\small{$\ddot\alpha$} \end{sideways}}
\end{overpic}
& 
\includegraphics[width=.45\textwidth]{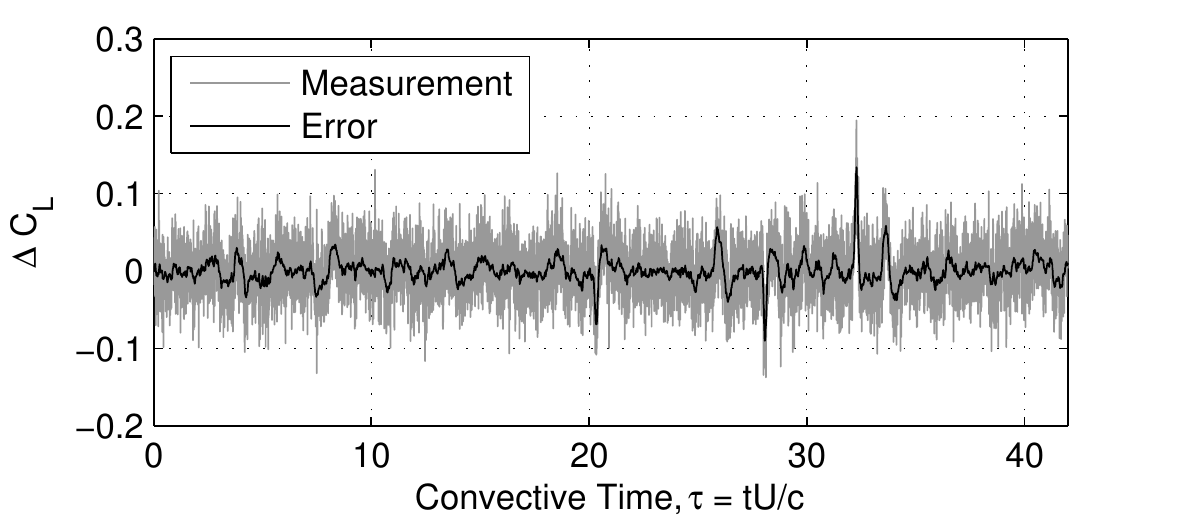} 
\end{tabular}
\vspace{-.15in}
\caption{Controller performance on aggressive maneuver with the addition of sensor noise.}\label{fig:controlleraggressive}
\end{center}
\vspace{-.2in}
\end{figure}

Figure~\ref{fig:controlleraggressive} shows the three model-based controllers on an aggressive reference lift tracking maneuver about the various base angles of attack, $\alpha_0$.  White noise, $n$, with 0.001 variance is added to the sensor; the noisy input to the controller, $e=r_L-C_L-n$, and corresponding noiseless error signal, $r_L-C_L$, are shown in the bottom right panel.  There are no disturbances added since the large angle of attack nonlinearity is disturbance enough.  All of the controllers track the maneuver quite well.  However, the angle of attack curve for the $\alpha_0=20^{\circ}$ maneuver deviates significantly from the other curves, indicating the nonlinearity excited at large angles, as well as differences in the controller.  The controller based at $\alpha_0=20^{\circ}$ has to work more to achieve the large unsteady lift coefficient.

%%%%%%%%%%%%%%%%%%%%%%%%
%%% SUMMARY
%%%%%%%%%%%%%%%%%%%%%%%%
\section{Discussion}\label{chap:models:summary}
This work demonstrates the use of low-order, state-space aerodynamic models, such as those previously developed in~\cite{brunton:2012b}, for robust closed-loop feedback control.  We first extend these models to include multiple inputs and explicit parameterization by the pitch-axis.  These models are state-space realizations of accurate indicial response models obtained from direct numerical simulations, and they capture the transient, viscous effects that are not modeled by inviscid theories.  In addition to extending these models, a careful methodology is presented to identify models from input--output data.  In particular, multiple-input, multiple-output extensions are discussed and case studies of the incorrect application of ERA/OKID are illustrated with likely pitfalls.  

The modeling and control results are demonstrated for a pitching and plunging flat plate at Reynolds number~100.  A robust feedback controller is developed to track reference lift coefficients while attenuating noise and compensating for disturbances.  Three controllers are constructed based on models linearized at $\alpha_0=0^\circ$, $\alpha_0=10^\circ$, and $\alpha_0=20^\circ$.  All three controllers perform well on a step-up in lift coefficient from $C_L(\alpha_0)$ to $C_L(\alpha_0)+0.5$, as well as on an aggressive maneuver with additive sensor noise.  In each case, the controller based at $\alpha_0=20^\circ$ must command larger relative angles of attack for the same lift, due to the decrease in lift coefficient slope $C_{\alpha}$.

The excellent controller performance, despite highly nonlinear separated flow effects, is due in large part to the instantaneous added-mass forces in response to accelerations of the plate.  For pitching between the leading-edge and mid-chord, these forces are favorable and have the same sign as the quasi-steady forces.  Right half plane zeros appear for pitching aft of the mid-chord, as the added-mass forces change sign.  In this case, a plate undergoing fast pitch-up will experience a brief dip in lift coefficient, due to negative added-mass, before the circulatory forces kick in and the lift rises.  This non-minimum-phase behavior, due to the appearance of right-half-plane zeros, complicates controller design and limits the controller bandwidth.  Developing controllers for pitching aft of the mid-chord, and investigating the consequent performance limitations is the subject of future work.

Finally, generalizing the methods in this work to include nonlinear flow phenomena will be an important contribution.  The goal is a nonlinear model that yields the models from this work when linearized about a specific angle of attack, yet has the correct nonlinear structure for large amplitude maneuvers such as in~\cite{noack:03cyl}.  In contrast to developing fully nonlinear models, gain-scheduled controllers are a plausible solution for control.

\section*{Acknowledgements}
The authors gratefully acknowledge the support for this work from the Air Force Office of Scientific Research grant FA9550-12-1-0075, and by the FAA, under the Joint University Program.  We thank Dave Williams and Wes Kerstens for valuable discussions on aerodynamic models.  We also thank Jeff Eldredge, Michael Ol, and the AIAA fluid dynamics technical committee's low-Reynolds number aerodynamics discussion group.

\setcounter{section}{0}
\renewcommand\thesection{\Alph{section}}

%%%%%%%%%%%%%%%%%%%%%%%%
%%% MANEUVERS
%%%%%%%%%%%%%%%%%%%%%%%%
\section{System identification maneuvers}\label{chap:models:maneuvers}
The maneuvers in this section are used in conjunction with the algorithms in Section~\ref{chap:models:algorithms} to develop models of the form in Section~\ref{chap:models:linear} from either numerical or experimental data~\cite{brunton:2012b}.  The algorithms in Sections~\ref{chap:models:algorithms:1} and~\ref{chap:models:algorithms:2} rely on the step response of $\alpha$ (or $\dot g, \dot h$).  A smoothed step function that is useful in simulations is discussed in Section~\ref{chap:models:maneuvers:step}.  Section~\ref{chap:models:maneuvers:okid} presents maneuvers that are used with the OKID method in Section~\ref{chap:models:algorithms:3} to obtain the impulse response in $\ddot\alpha$ (or $\ddot g,\ddot h$) from a more realistic input maneuver.  These aggressive pseudo-random maneuvers are particularly useful for experiments, since they excite various frequencies and overcome noisy measurements.

The measurements from a direct numerical simulation (DNS) of the Navier-Stokes (NS) equations or from a wind tunnel experiment are necessarily a discrete-time signal.  A fine timestep $\Delta t_f$ is required for the DNS to remain stable and approximate the continuous-time NS equations.  The transient aerodynamic effects, however, are modeled as a discrete-time system with a coarse timestep $\Delta t_c\gg \Delta t_f$.  Thus, we command maneuvers defined by a coarse discrete-time signal $u_k$, and simulate a corresponding smoothed discrete-time signal $\tilde{u}_{j}$ with timestep $\Delta t_f$.

% STEP FUNCTION
\subsection{Smoothed step functions}\label{chap:models:maneuvers:step}
For a number of reasons, an actual step response is non physical.  First, it is impossible to command in experiments or simulations, because it would correspond to a body instantaneously dematerializing and then rematerializing it in another location.  An alternative is to use a smoothed step maneuver and approach the limit as the maneuver becomes very rapid.  As the maneuver becomes increasingly rapid, the added-mass forces begin to dominate; in fact, a good rule of thumb is to choose a maneuver rapid enough that the lift response for the duration of the maneuver is dominated by added-mass forces.

The duration of the step maneuvers used for the results in this paper are either $\Delta t_c=0.01$ or $\Delta t_c=0.1$ convective time units.  The amplitude is either $M=0.1^{\circ}\approx 0.0017451~\text{rad}$ in the case of pitching or $M=0.0017451$ chord lengths per convection time in the case of vertical velocity, corresponding to $0.1^{\circ}$ change in effective angle of attack.  This is sufficiently rapid for the added-mass forces to dominate for the duration of the maneuver.  To obtain a model for plunging, we use a step-up in vertical velocity.

A pitch-up, hold, pitch-down maneuver was introduced by~\cite{eldredge:2009} as a canonical pitching maneuver to compare and study various experiments, simulations and models.  The linear ramp-step maneuver is based on the pitch-up, hold portion of the canonical maneuver, and the equations for $u$ and $\dot u$ are:
\begin{align}
u(t)&=M\frac{G(t)}{\max G(t)}& & \dot u(t)=M\frac{\tanh(b(t-t_1))-\tanh(b(t-t_2))}{\max G(t)}\label{eq:PRH3}
\end{align}
where
\begin{equation}
G(t) = \log\left[\frac{\cosh(b(t-t_1))}{\cosh(b(t-t_2))}\cdot\frac{\cosh(bt_2)}{\cosh(bt_1)}\right].
\label{eq:PRH1}
\end{equation}

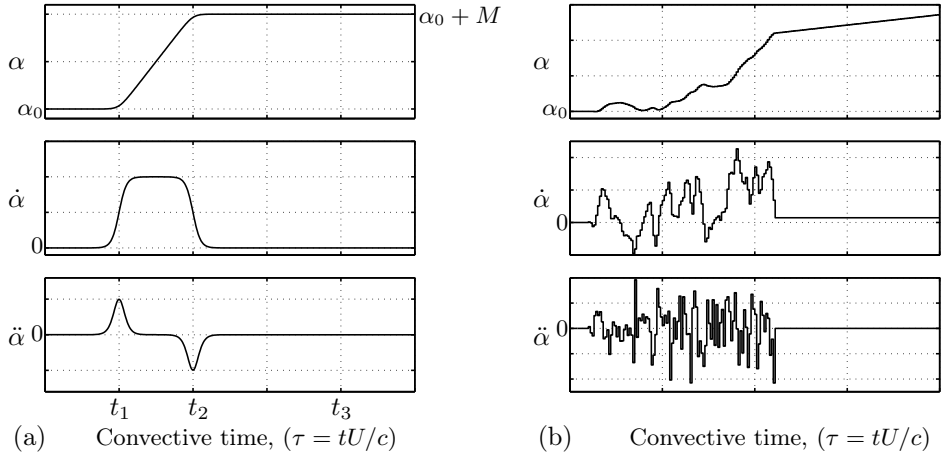
\begin{figure}
\begin{center}
\begin{tabular}{ccc}
\begin{overpic}[width=.35\textwidth]{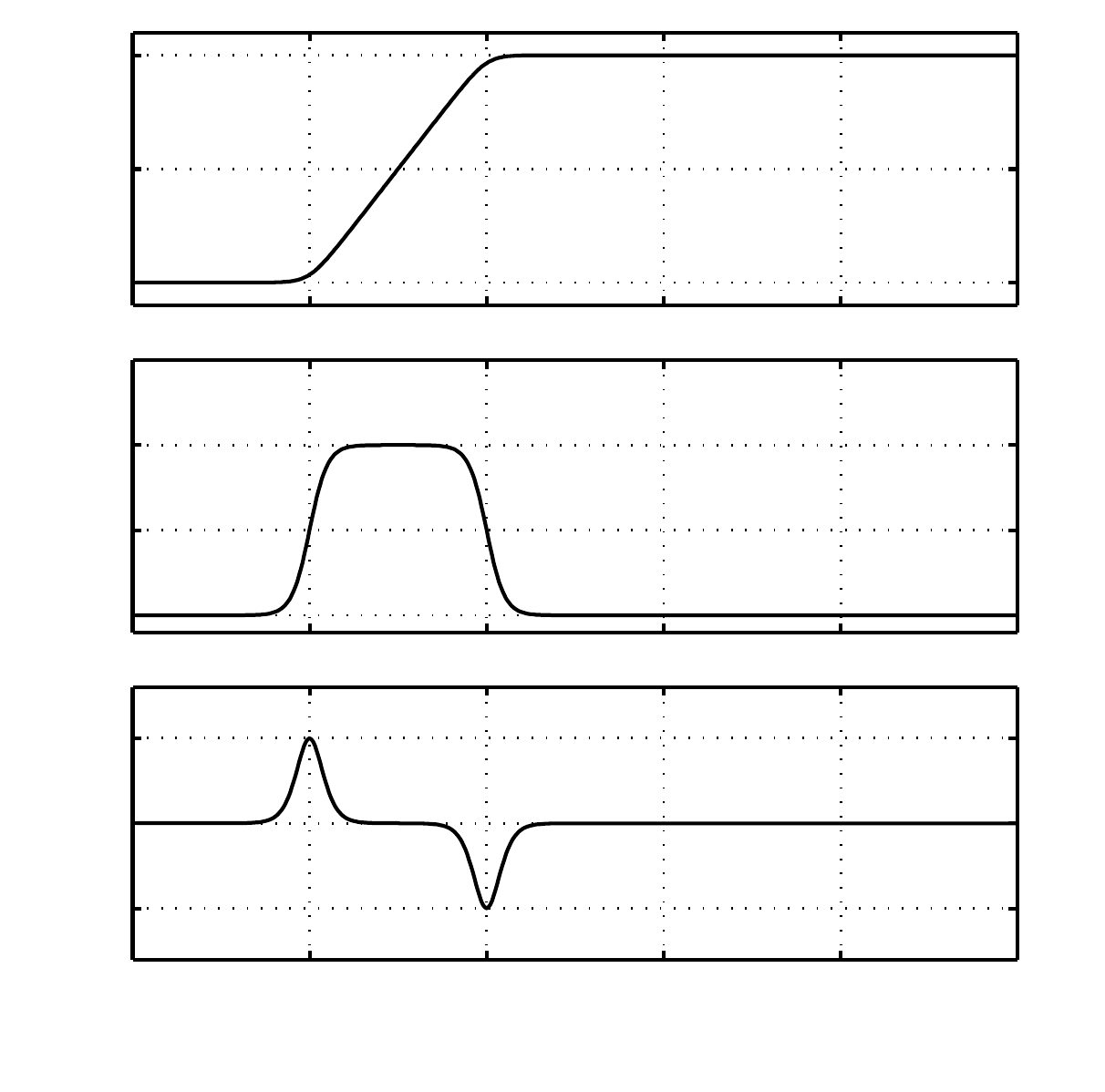}
\put(6,72){\small{$\alpha_0$}}
\put(93,92){\small{$\alpha_0+M$}}
\put(9,42){\small{$0$}}
\put(9,23){\small{$0$}}
\put(4,82){{$\alpha$}}
\put(4,52){{$\dot\alpha$}}
\put(4,22){{$\ddot\alpha$}}
\put(26,7){\normalsize{$t_1$}}
\put(43,7){\normalsize{$t_2$}}
\put(74,7){\normalsize{$t_3$}}
\put(23,0.5){{\small Convective time, $(\tau=tU/c)$}}
\put(5,.5){(a)}
\end{overpic}
&&
\begin{overpic}[width=.35\textwidth]{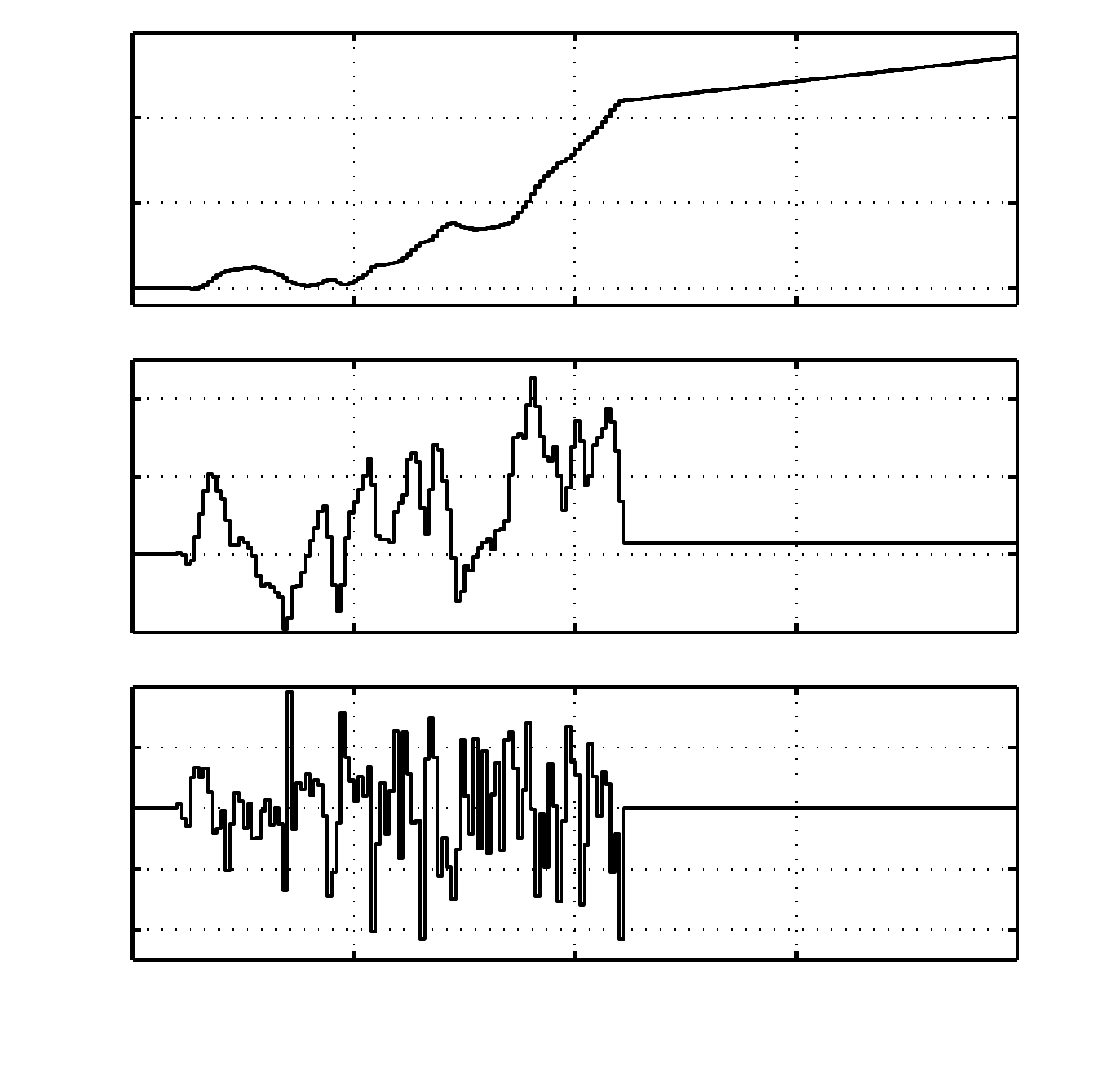}
\put(6,72){\small{$\alpha_0$}}
\put(9,47){\small{$0$}}
\put(9,24){\small{$0$}}
\put(4,82){{$\alpha$}}
\put(4,52){{$\dot\alpha$}}
\put(4,22){{$\ddot\alpha$}}
\put(23,0.5){{ \small Convective time, $(\tau=tU/c)$}}
\put(5,.5){(b)}
\end{overpic}
\end{tabular}
\caption{(a) Linear ramp in $u=\alpha$, approximating the discrete-time impulse $\dot\alpha_0=M/\Delta t_c$ for $\Delta t_c=t_2-t_1$.  $t_3$ is the duration of the ``hold" period for Section~\ref{chap:models:maneuvers:okid}, and it may be considered infinite for Section \ref{chap:models:maneuvers:step}.  (b)  Gaussian white noise input to $\ddot\alpha$.}
\label{fig:steps}
\vspace{-.15in}
\end{center}
\end{figure}

The maneuver is shown in Figure~\ref{fig:steps} (a) for a step in angle of attack, $u=\alpha$.  The start of the maneuver is $t=t_1$ and the duration of the ramp-up is $\Delta t_c=t_2-t_1$.  The parameter $b$ effects how gradual or abrupt the ramp acceleration is.  By choosing large $b$, it is possible to obtain a maneuver where the $\ddot{u}$ acceleration effects are localized near time $t_1=0$ and $t_2=\Delta t_c$, and the velocity $\dot{u}$ is constant throughout much of the maneuver.  This results in an approximately piecewise linear ramp-up, with smooth transitions at $t_1=0$ and $t_2=\Delta t_c$.

The linear ramp maneuver is a natural choice, since the boxy profile of the velocity $\dot{u}$ resembles a discrete-time impulse in $\dot{u}$ with timestep $\Delta t_c$.  Thus, it is possible to run simulations with a fine timestep $\Delta t_f$ that fully resolve the maneuver in time, and then down-sample to obtain a coarse discrete-time signal with sample time $\Delta t_c$.  

% OKID MANEUVERS
\subsection{Maneuvers for OKID method}\label{chap:models:maneuvers:okid}
The observer/Kalman filter identification (OKID) algorithm is useful for obtaining the Markov parameters (impulse response parameters) from more realistic input--output data.  For example, in an experiment, it may be difficult to obtain an impulse response in $\ddot\alpha$ because of the accompanying linear growth in $\alpha$.  Therefore, it is desirable to use a realistic maneuver with bounded $\alpha$ and estimate what the impulse response would be for the linearized system.  We seek models of the form in Eq.~(\ref{eq:rom3}) with $\ddot\alpha$ as the input, so the maneuvers must have rich frequency content in $\ddot\alpha$ to excite a large range of unsteady flow phenomena.  

Gaussian white noise may be used as the input to $\ddot\alpha$, as shown in Figure~\ref{fig:steps} (b).  The sample time for the white noise process is $\Delta t_c=0.1$ convective time.  This maneuver is primarily used with simulations, since the lift effect from various parts of the maneuver are subtle and may be overwhelmed by noise in an experiment. 

A system identification maneuver for wind tunnel experiments must satisfy a number of criteria.  Because the input to the model is $\ddot\alpha$, we ultimately need a maneuver with rich $\ddot\alpha$ content.  Also, because we are identifying various stability derivatives, $C_{\alpha}$, $C_{\dot\alpha}$, and $C_{\ddot\alpha}$, we need the contribution from individual changes in $\alpha, \dot\alpha$, and $\ddot\alpha$ to be distinguishable in the measured lift force.  Most importantly, obtaining high reduced frequencies in wind tunnel experiments typically involves lower tunnel velocity, which results in a low signal-to-noise ratio.  Therefore, the maneuver must be sufficiently aggressive so that the forces generated provide large signal to noise.

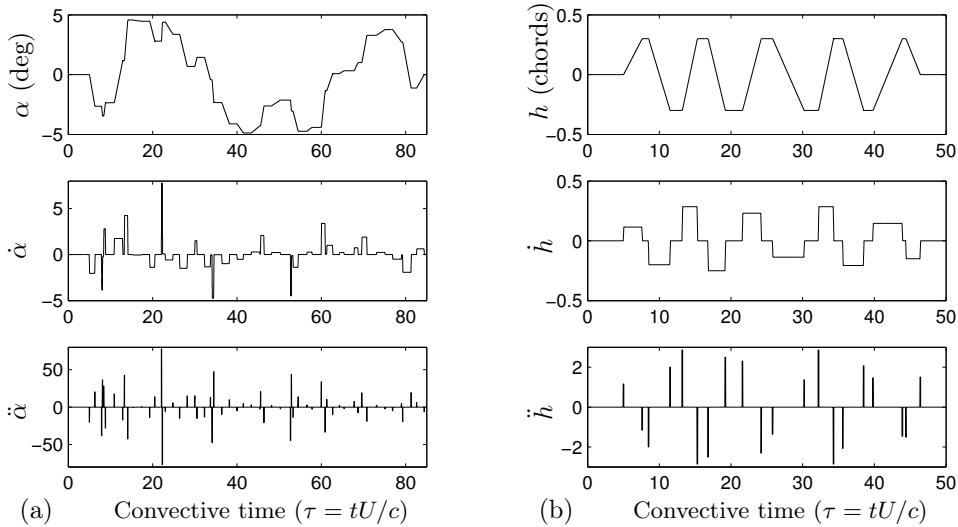
\begin{figure}
\begin{center}
\begin{tabular}{ccc}
\begin{overpic}[width=.35\textwidth]{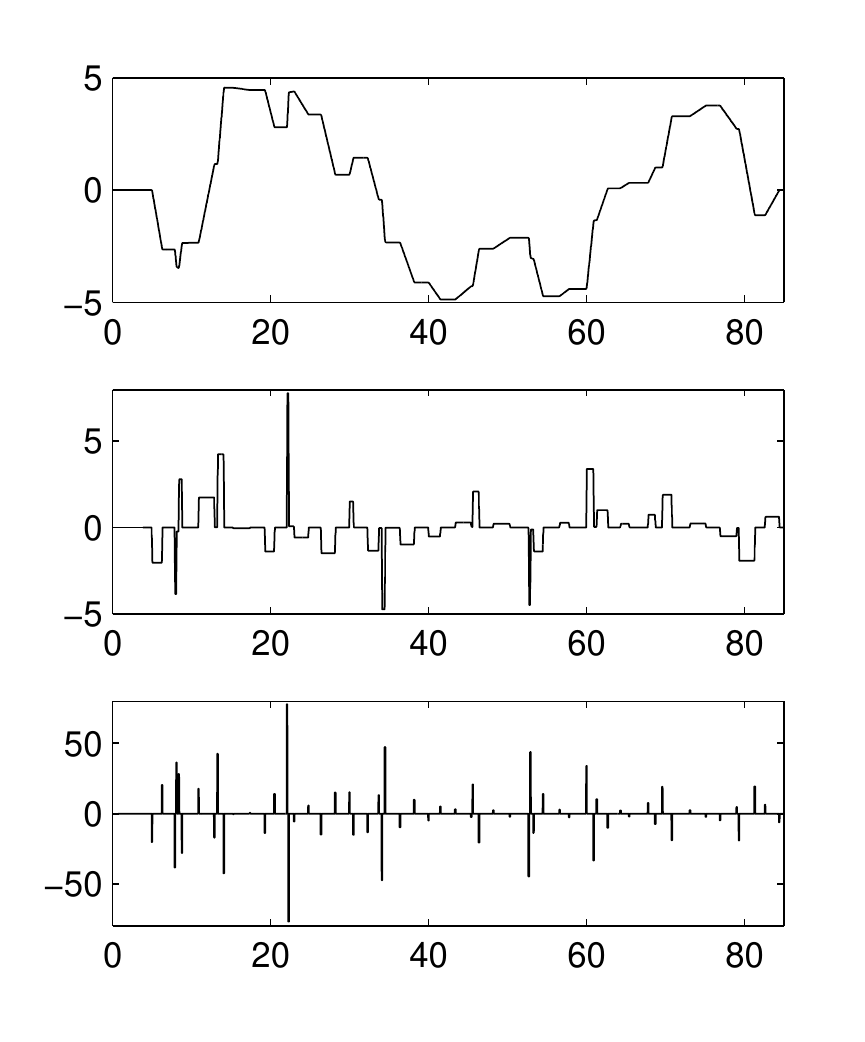}
\put(19,2){{\small Convective time ($\tau=tU/c$)}}
\put(0,19.5){\begin{sideways}$\ddot\alpha$\end{sideways}}
\put(0,48.5){\begin{sideways}$\dot\alpha$\end{sideways}}
\put(0,74){\begin{sideways}$\alpha$ (deg)\end{sideways}}
\put(2,2){(a)}
\end{overpic}&&
\begin{overpic}[width=.35\textwidth]{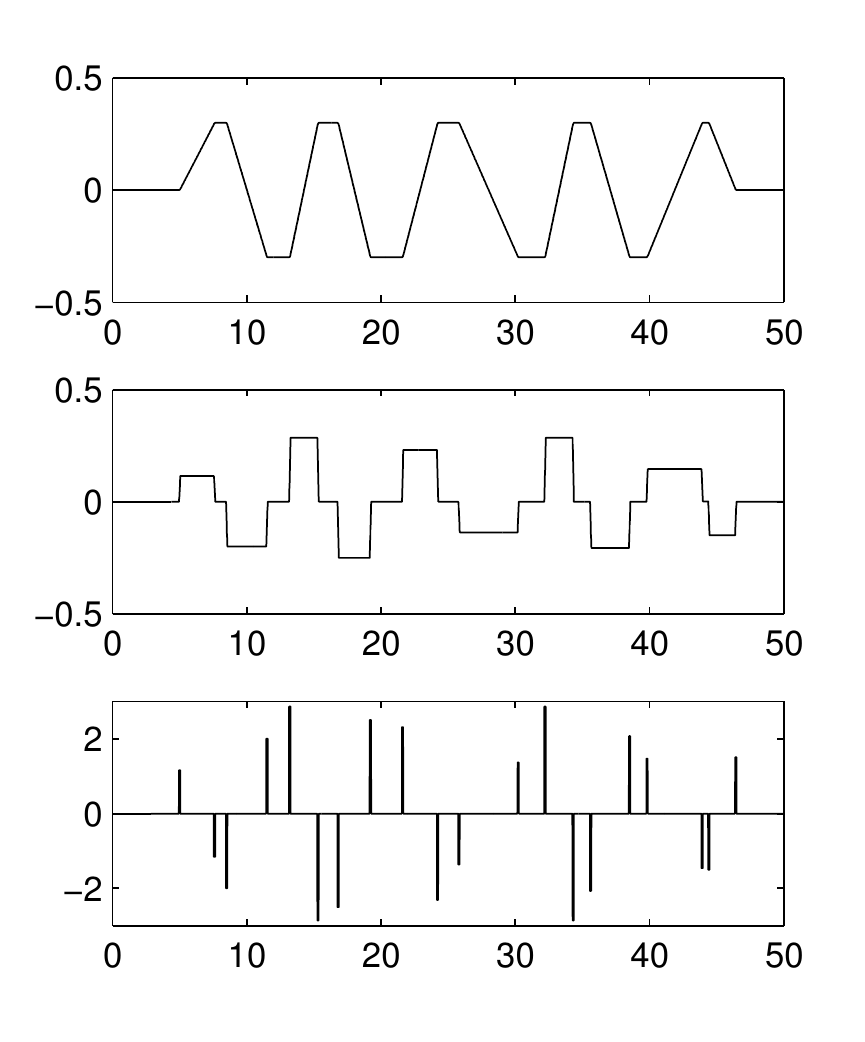}
\put(19,2){{\small Convective time ($\tau=tU/c$)}}
\put(0,19.5){\begin{sideways}$\ddot h$\end{sideways}}
\put(0,49.5){\begin{sideways}$\dot h$  \end{sideways}}
\put(0,73.5){\begin{sideways}$ h$ (chords) \end{sideways}}
\put(2,2){(b)}
\end{overpic}
\end{tabular}
\end{center}
\vskip -.2in
\caption{Aggressive maneuvers for identifying (a) pitch dynamics and (b) plunge dynamics.}
\label{fig:pitchmaneuver}
\vspace{-.2in}
\end{figure}

The maneuver, shown in Figure~{\ref{fig:pitchmaneuver} (a), is constructed as a pseudo-random sequence of ramp-up, hold and ramp-down, hold maneuvers, from Section~\ref{chap:models:maneuvers:step}, which are related to the canonical maneuvers in~\cite{eldredge:2009}.  The equations for $u$ and $\dot{u}$ for a single pitch-up, hold are given in Eq.~(\ref{eq:PRH3}), with $u=\alpha$.  

The duration of the ramp $\tau_r=t_2-t_1$ and hold $\tau_h=t_3-t_2$ are bounded Gaussian white noise processes.  The step amplitude $M$ is also sampled from a Gaussian distribution, with the constraint that the maneuver amplitude never exceeds $\pm 10^{\circ}$.  This maneuver is attractive because $\ddot\alpha$ consists of pseudo-randomly spaced pulses at the beginning and end of each ramp; the pseudo-random pulses are inspired by the work of \cite{Kerstens:2011}.  The result is that the large added-mass forces are similarly spaced pulses.  The large ramp-up and ramp-down maneuvers result in forces that are large compared with sensor noise.  A similar maneuver is generated for plunging motion, shown in Figure~\ref{fig:pitchmaneuver} (b).  Because there is no steady-state lift associated with a fixed vertical position $h$, it is unnecessary for the maneuver to sample different vertical positions during the hold periods, in contrast to the pitching maneuver.  

\bibliographystyle{plain}
\bibliography{MODELS}

\begin{thebibliography}{10}

\bibitem{ahuja:2010}
S.~Ahuja and C.~W. Rowley.
\newblock Feedback control of unstable steady states of flow past a flat plate
  using reduced-order estimators.
\newblock {\em Journal of Fluid Mechanics}, 645:447--478, 2010.

\bibitem{Allen:2001}
J.~J. Allen and A.~J. Smits.
\newblock Energy harvesting eel.
\newblock {\em Journal of Fluids and Structures}, 15:629--640, 2001.

\bibitem{Farhat:2010}
D.~Amsallem, J.~Cortial, and C.~Farhat.
\newblock Towards real-time computational-fluid-dynamics-based aeroelastic
  computations using a database of reduced-order information.
\newblock {\em AIAA Journal}, 48(9):2029--2037, 2010.

\bibitem{bagheri:2009}
S.~Bagheri, L.~Brandt, and D.~S. Henningson.
\newblock Input-output analysis, model reduction and control of the flat-plate
  boundary layer.
\newblock {\em Journal of Fluid Mechanics}, 620:263--298, 2009.

\bibitem{Balajewicz:2012}
M.~Balajewicz and E.~Dowell.
\newblock Reduced-order modeling of flutter and limit-cycle oscillations using
  the sparse {Volterra} series.
\newblock {\em Journal of Aircraft}, 49(6):1803--1812, 2012.

\bibitem{birch:01buglev}
J.~Birch and M.~Dickinson.
\newblock Spanwise flow and the attachment of the leading-edge vortex on insect
  wings.
\newblock {\em Nature}, 412:729--733, 2001.

\bibitem{Bruno:2008}
L.~Bruno and D.~Fransos.
\newblock Evaluation of {Reynolds} number effects on flutter derivatives of a
  flat plate by means of a computational approach.
\newblock {\em Journal of Fluids and Structures}, 24:1058--1076, 2008.

\bibitem{brunton:2012a}
S.~L. Brunton and C.~W. Rowley.
\newblock Empirical state-space representations for {T}heodorsen's lift model.
\newblock {\em Journal of Fluids and Structures}, 38:174--186, 2013.

\bibitem{brunton:2012b}
S.~L. Brunton, C.~W. Rowley, and D.~R. Williams.
\newblock Reduced-order unsteady aerodynamic models at low {R}eynolds numbers.
\newblock {\em Journal of Fluid Mechanics}, 724:203--233, 2013.

\bibitem{clark:2006}
R.~P. Clark and A.~J. Smits.
\newblock Thrust production and wake structure of a batoid-inspired oscillating
  fin.
\newblock {\em Journal of Fluid Mechanics}, 562:415--429, 2006.

\bibitem{taira:fastIBPM}
T.~Colonius and K.~Taira.
\newblock A fast immersed boundary method using a nullspace approach and
  multi-domain far-field boundary conditions.
\newblock {\em Computer Methods in Applied Mechanics and Engineering},
  197:2131--2146, 2008.

\bibitem{Combes:2001}
S.~A. Combes and T.~L. Daniel.
\newblock Shape, flapping and flexion: wing and fin design for forward flight.
\newblock {\em The Journal of Experimental Biology}, 204:2073--2085, 2001.

\bibitem{Costa:2007}
C.~Costa.
\newblock Aerodynamic admittance functions and buffeting forces for bridges via
  indicial functions.
\newblock {\em Journal of Fluids and Structures}, 23:413--428, 2007.

\bibitem{Dabiri:2009}
J.~O. Dabiri.
\newblock Optimal vortex formation as a unifying principle in biological
  propulsion.
\newblock {\em Annual Review of Fluid Mechanics}, 41:17--33, 2009.

\bibitem{Daniel:1984}
T.~L. Daniel.
\newblock Unsteady aspects of aquatic locomotion.
\newblock {\em American Zoologist}, 24(1):121--134, 1984.

\bibitem{dowell:2001}
E.~H. Dowell and K.~C. Hall.
\newblock Modeling of fluid-structure interaction.
\newblock {\em Annual Review of Fluid Mechanics}, 33:445--490, 2001.

\bibitem{eldredge:2009}
J.~D. Eldredge, C.~Wang, and M.~V. OL.
\newblock A computational study of a canonical pitch-up, pitch-down wing
  maneuver.
\newblock AIAA Paper 2009-3687, 39th Fluid Dynamics Conference, June 2009.

\bibitem{kalman:1965}
B.~L. Ho and R.~E. Kalman.
\newblock Effective construction of linear state-variable models from
  input/output data.
\newblock In {\em Proceedings of the 3rd Annual Allerton Conference on Circuit
  and System Theory}, pages 449--459, 1965.

\bibitem{ilak:2008}
M.~Ilak and C.~W. Rowley.
\newblock Modeling of transitional channel flow using balanced proper
  orthogonal decomposition.
\newblock {\em Physics of Fluids}, 20:034103, 2008.

\bibitem{illingworth:2010}
S.~J. Illingworth, A.~S. Morgans, and C.~W. Rowley.
\newblock Feedback control of flow resonances using balanced reduced-order
  models.
\newblock {\em Journal of Sound and Vibration}, 330(8):1567--1581, 2010.

\bibitem{ERA:1985}
J.~N. Juang and R.~S. Pappa.
\newblock An eigensystem realization algorithm for modal parameter
  identification and model reduction.
\newblock {\em Journal of Guidance, Control and Dynamics}, 8(5):620--627, 1985.

\bibitem{juang:1991}
J.~N. Juang, M.~Phan, L.~G. Horta, and R.~W. Longman.
\newblock Identification of observer/{Kalman} filter {Markov} parameters:
  Theory and experiments.
\newblock Technical Memorandum 104069, NASA, 1991.

\bibitem{Kerstens:2011}
W.~Kerstens, J.~Pfeiffer, D.~Williams, R.~King, and T.~Colonius.
\newblock Closed-loop control of lift for longitudinal gust suppression at low
  {R}eynolds numbers.
\newblock {\em AIAA Journal}, 49(8):1721--1728, 2011.

\bibitem{Leishman:1994}
J.~G. Leishman.
\newblock Unsteady lift of a flapped airfoil by indicial concepts.
\newblock {\em Journal of Aircraft}, 31(2):288--297, 1994.

\bibitem{Leishman:1996}
J.~G. Leishman.
\newblock Subsonic unsteady aerodynamics caused by gusts using the indicial
  method.
\newblock {\em Journal of Aircraft}, 33(5):869--879, 1996.

\bibitem{leishman:06}
J.~G. Leishman.
\newblock {\em Principles of Helicopter Aerodynamics}.
\newblock Cambridge University Press, Cambridge, England, 2 edition, 2006.

\bibitem{ERA:2009}
Z.~Ma, S.~Ahuja, and C.~W. Rowley.
\newblock Reduced order models for control of fluids using the eigensystem
  realization algorithm.
\newblock {\em Theoretical and Computational Fluid Dynamics}, 25(1):233--247,
  2011.

\bibitem{Marzocca:2002}
P.~Marzocca, L.~Librescu, and W.~A. Silva.
\newblock Aeroelastic response of nonlinear wing sections using a functional
  series technique.
\newblock {\em AIAA Journal}, 40(5):813--824, 2002.

\bibitem{noack:03cyl}
B.~R. Noack, K.~Afanasiev, M.~Morzynski, G.~Tadmor, and F.~Thiele.
\newblock A hierarchy of low-dimensional models for the transient and
  post-transient cylinder wake.
\newblock {\em Journal of Fluid Mechanics}, 497:335--363, 2003.

\bibitem{canonical:2010}
M.~V. OL, A.~Altman, J.~D. Eldredge, D.~J. Garmann, and Y.~Lian.
\newblock R\'esum\'e of the {AIAA} {FDTC} low {Reynolds} number discussion
  group's canaonical cases.
\newblock AIAA Paper 2010-1085, 48th Aerospace Sciences Meeting, January 2010.

\bibitem{Prazenica:2007}
R.~J. Prazenica, P.~H. Reisenthel, A.~J. Kurdila, and M.~J. Brenner.
\newblock Volterra kernel extrapolation for modeling nonlinear aeroelastic
  systems at novel flight conditions.
\newblock {\em Journal of Aircraft}, 44(1):149--162, 2007.

\bibitem{rowley:05pod}
C.~W. Rowley.
\newblock Model reduction for fluids using balanced proper orthogonal
  decomposition.
\newblock {\em International Journal of Bifurcation and Chaos},
  15(3):997--1013, 2005.

\bibitem{salvatori:2006}
L.~Salvatori and P.~Spinelli.
\newblock Effects of structural nonlinearity and along-span wind coherence on
  suspension bridge aerodynamics: Some numerical simulation results.
\newblock {\em Journal of Wind Engineering and Industrial Aerodynamics},
  94:415--430, 2006.

\bibitem{sane:03bug}
S.~P. Sane.
\newblock The aerodynamics of insect flight.
\newblock {\em The Journal of Experimental Biology}, 206(23):4191--4208, 2003.

\bibitem{Shelley:2011}
M.~J. Shelley and J.~Zhang.
\newblock Flapping and bending bodies interacting with fluid flows.
\newblock {\em Annual Review of Fluid Mechanics}, 43:449--465, 2011.

\bibitem{Silva:2004}
W.~A. Silva and R.~E. Bartels.
\newblock Development of reduced-order models for aeroelastic analysis and
  flutter prediction using the {CFL3Dv6.0} code.
\newblock {\em Journal of Fluids and Structures}, 19:729--745, 2004.

\bibitem{sp:book}
S.~Skogestad and I.~Postlethwaite.
\newblock {\em Multivariable Feedback Control: Analysis and Design}.
\newblock John Wiley \& Sons, Inc., Hoboken, New Jersey, 2 edition, 2005.

\bibitem{taira:07ibfs}
K.~Taira and T.~Colonius.
\newblock The immersed boundary method: a projection approach.
\newblock {\em Journal of Computational Physics}, 225(2):2118--2137, 2007.

\bibitem{theodorsen:35}
T.~Theodorsen.
\newblock General theory of aerodynamic instability and the mechanism of
  flutter.
\newblock Technical Report 496, NACA, 1935.

\bibitem{tobak:1954}
M.~Tobak.
\newblock On the use of the indicial function concept in the analysis of
  unsteady motions of wings and wing-tail combinations.
\newblock Report 1188, NACA, 1954.

\bibitem{tobak:85}
M.~Tobak and G.~T. Chapman.
\newblock Nonlinear problems in flight dynamics involving aerodynamic
  bifurcations.
\newblock Technical Memorandum 86706, NASA, 1985.

\bibitem{valasek:2003}
J.~Valasek and W.~Chen.
\newblock Observer/{Kalman} filter identification for online system
  identification of aircraft.
\newblock {\em Journal of Guidance, Control and Dynamics}, 26(2):347--353,
  2003.

\bibitem{wagner:25}
H.~Wagner.
\newblock {\"Uber} die {Entstehung} des dynamischen {Auftriebes} von
  {Tragfl\"ugeln}.
\newblock {\em Zeitschrift f\"ur Angewandte Mathematic und Mechanik},
  5(1):17--35, 1925.

\bibitem{wang:2005}
Z.~J. Wang.
\newblock Dissecting insect flight.
\newblock {\em Annual Review of Fluid Mechanics}, 37:183--210, 2005.

\bibitem{Wu:2011}
T.~Y. Wu.
\newblock Fish swimming and bird/insect flight.
\newblock {\em Annual Review of Fluid Mechanics}, 43:25--58, 2011.

\end{thebibliography}
\end{document}